\documentclass[manuscript,screen,nonacm]{acmart}

\AtBeginDocument{%
  \providecommand\BibTeX{{%
    \normalfont B\kern-0.5em{\scshape i\kern-0.25em b}\kern-0.8em\TeX}}}

\copyrightyear{2024}
\setcopyright{acmcopyright}
\acmYear{2024}
\acmDOI{XXXXXXX.XXXXXXX}

\acmConference[WSDM '24]{
The 17th ACM International Conference on Web Search and Data Mining
}{March 4--8,
  2024}{Mérida, México}
\acmBooktitle{
Proceedings of the Seventeenth ACM International Conference on Web Search and Data Mining (WSDM ’24), March 4--8, 2024, Mérida, México
} 
\acmPrice{15.00}
\acmISBN{978-1-4503-XXXX-X/18/06}

\usepackage[utf8]{inputenc}

\usepackage{url}
\usepackage{amsmath}
\usepackage{amsthm}

\usepackage{microtype}
\usepackage{subfigure}
\usepackage{booktabs} 
\usepackage{amsmath}
\usepackage{amsthm}
\usepackage{amsfonts}       
\usepackage{mathtools}
\usepackage{nicefrac}
\usepackage{algorithm,algorithmic}
\usepackage{listings}
\usepackage{balance}

\usepackage{enumerate}
\usepackage{enumitem}

\usepackage{latexsym}

\newcommand{\commentout}[1]{}

\newcommand{\bfT}{\mathbf{T}}

\newcommand{\bfD}{\mathbf{D}}

\newcommand{\bfv}{\mathbf{v}}
\newcommand{\bfw}{\mathbf{w}}

\newcommand{\bfR}{\mathbf{R}}

\newcommand{\calH}{\mathcal{H}}
\newcommand{\calI}{\mathcal{I}}
\newcommand{\calL}{\mathcal{L}}
\newcommand{\calN}{\mathcal{N}}

\newcommand{\calQ}{\mathcal{Q}}
\newcommand{\calR}{\mathcal{R}}
\newcommand{\calT}{\mathcal{T}}
\newcommand{\calU}{\mathcal{U}}
\newcommand{\calX}{\mathcal{X}}

\newcommand{\olg}{\overline{g}}

\DeclareMathOperator*{\argmax}{argmax}

\newcommand{\veps}{\varepsilon}

\newcommand{\NumR}{\mathit{Num}}
\newcommand{\Rated}{\mathit{Rated}}
\newcommand{\Tagged}{\mathit{Tagged}}
\newcommand{\PT}{\mathit{PT}}

\newcommand{\EVOI}{\mathit{EVOI}}
\newcommand{\PEU}{\mathit{PEU}}
\newcommand{\EU}{\mathit{EU}}
\newcommand{\IG}{\mathit{IG}}
\newcommand{\RQ}{\mathit{RQ}}

\newtheoremstyle{TheoremNum}%
    {\topsep}{\topsep}
    {\itshape}
    {}
    {\bfseries}
    {.}
    { }
    {\thmname{#1}\thmnote{ \bfseries #3}}
\theoremstyle{TheoremNum}

\newcommand{\EE}{\mathbb{E}}

\def\I{\mathcal{I}}

\def\N{\mathcal{N}}

\usepackage[textwidth=18mm]{todonotes}

\begin{document}

\title{Preference Elicitation with Soft Attributes in Interactive Recommendation}


\author{Erdem B{\i}y{\i}k}
\affiliation{%
  \institution{Thomas Lord Department of Computer Science, University of Southern California}
  \streetaddress{3737 Watt Way}
  \city{Los Angeles}
  \country{USA}
  \postcode{90089}
}
\email{biyik@usc.edu}

\author{Fan Yao}
\affiliation{%
  \institution{University of Virginia}
  \streetaddress{1827 University Avenue}
  \city{Charlottesville}
  \state{VA}
  \country{USA}
  \postcode{22903}
}
\email{fy4bc@virginia.edu}

\author{Yinlam Chow}
\authornote{Contact author.}
\affiliation{%
  \institution{Google Research}
  \streetaddress{1600 Amphitheatre Parkway}
  \city{Mountain View}
  \state{CA}
  \country{USA}
  \postcode{94043}
}
\email{yinlamchow@google.com}

\author{Alex Haig}
\affiliation{%
  \institution{Google Research}
  \streetaddress{1600 Amphitheatre Parkway}
  \city{Mountain View}
  \state{CA}
  \country{USA}
  \postcode{94043}
}
\email{ahaig@google.com}

\author{Chih-wei Hsu}
\affiliation{%
  \institution{Google Research}
  \streetaddress{1600 Amphitheatre Parkway}
  \city{Mountain View}
  \state{CA}
  \country{USA}
  \postcode{94043}
}
\email{cwhsu@google.com}

\author{Mohammad Ghavamzadeh}
\affiliation{%
  \institution{Google Research}
  \streetaddress{1600 Amphitheatre Parkway}
  \city{Mountain View}
  \state{CA}
  \country{USA}
  \postcode{94043}
}
\email{ghavamza@google.com}

\author{Craig Boutilier}
\affiliation{%
  \institution{Google Research}
  \streetaddress{1600 Amphitheatre Parkway}
  \city{Mountain View}
  \state{CA}
  \country{USA}
  \postcode{94043}
}
\email{cboutilier@google.com}

\renewcommand{\shortauthors}{B{\i}y{\i}k, Yao, Chow, Hsu, and Haig et al.}

\begin{abstract}

\emph{Preference elicitation} plays a central role in interactive \emph{recommender systems}. Most preference elicitation approaches use either item queries that ask users to select preferred items from a slate, or attribute queries that ask them to express their preferences for item characteristics. Unfortunately, users often wish to describe their preferences using \emph{soft} attributes \citep{sigir21:filipandkristian} for which no ground-truth semantics is given. Leveraging \emph{concept activation vectors} \citep{gopfert2021discovering} for soft attribute semantics, we develop novel preference elicitation methods that can accommodate soft attributes and bring together both item and attribute-based preference elicitation. Our techniques query users using both items and soft attributes to update the recommender system's belief about their preferences to improve recommendation quality. We demonstrate the effectiveness of our methods \emph{vis-\`{a}-vis} competing approaches on both synthetic and real-world datasets.
\end{abstract}


\maketitle

\section{Introduction}
\label{sec:intro}

Recommender systems (RSs) play a central role in connecting users to products, content and services by predicting user \emph{preferences} for candidate items. While practical RSs are often trained using implicit user feedback on recommended items (e.g., clicks, consumption, ratings), increasingly interactive and conversational systems explicitly elicit user preferences to more effectively guide recommendations. Approaches to \emph{preference elicitation (PE)} include both item-based \citep{Sarwar:www01, cf-survey:2005}
and attribute-based methods \citep{survey_pref_elicit:04,viappiani:06}, 
but rarely are these considered in conjunction. Moreover, attribute-based methods generally rely on \emph{hard attributes}, those for which a definitive source of ground truth associates attribute values (e.g., actors in movies, color of a product) with items.
Unfortunately, attributes with which users often wish to describe their preferences are \emph{soft} \citep{sigir21:filipandkristian}---there is 
 no source of ground truth associating such attribute values to items; instead this association, or \emph{semantics}, must be learned.  For instance, 
 information about movie attributes like `funny,' `thought-provoking,' or `inspiring' may only be learnable via sparse, noisy user comments, reviews, or tags. Because of this, practical attribute-based PE must account for the inherent noise and uncertainty in the semantics of soft attributes.\footnote{Soft attribute usage may also be \emph{subjective} \citep{gopfert2021discovering}, an issue we set aside in this work.}
 
In this work, we develop PE methods that can accommodate such soft attributes in item-based and attribute-based PE. To do so, we leverage the recent method by ~\citet{gopfert2021discovering}, who use \emph{concept activation vectors (CAVs)} \citep{kimTCAV:icml18} to discover the semantics of soft attributes w.r.t.\ a RS's item representation. We show how to incorporate this semantics into interactive RSs with \emph{active critiquing}: given a slate of items, our techniques determine the soft attributes about which to elicit user feedback (e.g., ``do you prefer a funnier or a less funny movie than this slate of movies?''), interpret the user's response relative to this semantics, and update RS beliefs about the user's preferences. 
Because the semantics are learned from sparse and noisy data, the RS may have varying degrees of certainty or confidence in the CAV semantics of different soft attributes.
This should, in turn, impact how the RS interprets user responses to attribute-based PE-queries and updates its beliefs. We extend the CAV semantics to handle such uncertainty and incorporate this into PE-driven belief updates.
 %

%
Our key contributions are as follows: (i) we leverage CAV semantics to extend attribute-based PE for interactive RSs to handle soft attributes (in contrast to \citet{gopfert2021discovering}, who model user-initiated critiques); (ii) we propose two novel query types: queries over attributes, and their combination with well-studied item queries, along with the human response models to these new query types, which enable us to efficiently update RS beliefs over user preferences in our PE framework; (iii) we propose several PE (query) selection schemes and optimization methods that balance information gathering and recommendation quality; and (iv) we extend our techniques to incorporate uncertainty in an attribute's CAV semantics.
We illustrate the efficacy of our methods on both synthetic data and MovieLens data \citep{harper16:movielens}.


\section{Problem Formulation}
\label{sec:formulation}

In this section, we outline our problem formulation and key assumptions, then briefly discuss related work.


\noindent \textbf{2.1 Preference Prediction.}
We assume a standard \emph{collaborative filtering (CF)} setting \citep{Hu:2008}, where users $u\in\calU$ rate items $i\in\calI$ with $r_{u,i}\in\calR\cup\{0\}$. Here $\calR$ is the set of possible ratings (e.g.,~$1$--$5$ stars) and $r_{u,i}=0$ indicates that user $u$ has not rated item $i$. Let the {\em ratings dataset} be $\bfR = \{(u,i,r_{u,i}) : r_{u,i} \neq 0\}$. 

The RS learns user and item representations from the ratings dataset $\bfR$
using some form of CF. Let $\calX\subseteq \mathbb R^d$ be a user-item embedding space. An \emph{item embedding} $\phi_I:\calI\rightarrow\calX$ maps each item $i\in\calI$ into a vector representation $\phi_I(i)$ of its (latent) attributes in $\calX$; note that this embedding is typically not interpretable. We assume that user preferences are captured by a similar mapping $\phi_U:\calU\rightarrow\calX$. However, to capture  the preference variations of users due to their latent state (e.g., mood, activity, location), the RS learns an \emph{embedding distribution} for each user, i.e.,~$P_U:\calU\rightarrow \mathbb{P}(\calX)$. As such, the embedding $\phi_U(u) = \phi_u\sim P_U(u)$ in $u$'s session is a sample from her embedding distribution. Since  $\phi_u$ is unknown to the RS, we use PE to uncover it.\footnote{Our techniques can be applied, \emph{mutatis mutandis}, to the RS's uncertainty about a user's \emph{stationary preferences} and its gradual refinement of beliefs over time, rather than assuming that preferences actually change (are resampled) with each session.}


To learn user and item embeddings, we use a \emph{two-tower model} (or \emph{dual encoder}) in which users and items are passed through separate but co-trained deep neural networks (DNNs) \citep{yiEtAl:recsys19,yangEtAl:www20}. The item encoder outputs $\phi_I(i)$ for any item $i\in\calI$, while the user encoder outputs a $d$-dimensional multivariate Gaussian distribution $P_U(u) =\mathcal N(\phi_{\mu,U}(u), \phi_{\sigma,U}(u))$ for any user $u\in\calU$, where $\phi_{\mu,U}(u)$ is
the mean user embedding vector and $\phi_{\sigma,U}(u)$ is a (lower-triangular) scale matrix. Suitable methods 
include probabilistic matrix factorization~\citep{salakhutdinov-mnih:nips07} and certain forms of neural CF~\citep{beutel_etal:wsdm18,yangEtAl:www20}. These methods often assume linear user utility, i.e., (predicted) user-item affinity is $\hat{r}_{u,i} = \phi_u^\top \phi_I(i)$. Hence, we treat $\phi_u$ as the parameters of $u$'s utility function (w.r.t.\ the learned item representation).
Our Gaussian user embedding gives the following \emph{belief state} over the user's utility function: $P(r_{u,i} \mid P_U, \phi_I)=\mathcal N( \phi_{\mu,U}(u)^\top \phi_I(i), \phi_I(i)^\top\phi_{\sigma,U}(u)^\top\phi_{\sigma,U}(u) \phi_I(i))$. 
We train these encoders by minimizing negative log-likelihood:
$\mathcal{L} = -\frac{1}{|\bfR|} \sum_{(u,i,r_{u,i})\in\bfR}\log P(r_{i,u} \mid P_U, \phi_I)$.

\vskip 1mm
\noindent
\textbf{2.2 \hskip 1mm Soft Attributes \& Tags.}
In addition to \emph{hard (known or objective) attributes}, users often describe items using \emph{soft attributes} which have no ``ground truth'' semantics or association with items. 
Unlike hard attributes, such as `genre,' `artist,' or `director,' these terms are not part of an agreed-upon formal specification of an item. They are neither applied universally to all items, nor by all users.

Some RSs support user-supplied {\em tags} (e.g., MovieLens), which may be specified directly in the RS, or extracted from user descriptions, reviews, or other sources. A tag can denote a hard or soft attribute. We let $\calT$ be the set of tags that users may adopt to describe items.
We assume that tags are used in a binary way---users simply choose to apply a tag or not. However, the actual attributes corresponding to tags can be ordinal or cardinal.
Indeed, the user study in \citet{sigir21:filipandkristian} shows that soft attributes often exhibit ``relative'' degrees.
For example, applying the tag `violent' may reflect that a user believes a movie exhibits a \emph{degree of violence} exceeding her tolerance.
Let $t_{u,i,g}=1$ if user $u$ applies tag $g\in\calT$ to item $i$, and $t_{u,i,g}=0$ otherwise. 
For each $g\in\calT$, let $\bfT_g = \{(u,i) : t_{u,i,g}= 1\}$ and $\bfT_{\olg} = \{(u,i) : t_{u,i,g}= 0, \; \exists g'\in\calT: t_{u,i,g'}= 1, \; \exists i'\in\calI : t_{u,i',g}=1\}$. Pair $(u,i)$ belongs to $\bfT_g$, if user $u$ applies tag $g$ to item $i$, and belongs to $\bfT_{\olg}$, if $u$ applies a tag $g'\neq g$ to $i$ and tags another item $i'\neq i$ with $g$.
Tags are usually strictly sparser than ratings, so we assume
$\sum_g |\bfT_g| \ll |\bfR|$. 

\vskip 1mm
\noindent
\textbf{2.3 \hskip 1mm Concept Activation Vectors in RSs.} 
Modern ML models usually learn complex and non-transparent representations of concepts, an issue addressed by work on interpretable representations~\citep{sundararajan:icml2017}. \citet{kimTCAV:icml18} develop an approach that tries to find a correspondence between internal state of an ML model (e.g., a DNN used for image classification)---in the form of a \emph{concept activation vector (CAV)}---and a human-interpretable concept.

\citet{gopfert2021discovering} apply CAVs to identify the semantics of tags w.r.t.\ the item representation learned by CF models in RSs. We adopt this approach in our work: for a tag $g\in\calT$, we attempt to find a CAV $\phi_g\in \calX$ in the embedding space, and use it to determine if the CF model has learned a representation of (the attribute corresponding to) $g$.\footnote{While a tag $g$ is simply a \emph{token} used by users to label items, its corresponding soft attribute is a \emph{property} of the items themselves, which can take a real-value. For conciseness, we sometimes refer to the semantics (or representation or CAV) of a tag $g$ as shorthand for the semantics of the attribute corresponding to $g$.} We train the CF model and learn CAVs separately, similar to methods that build attribute models on top of embeddings for cold-start \citep{rendle2010,cohen:recsys2017}, and in contrast to those that jointly train attribute models \citep{wu2019deep,luo2020latent}. This separation is motivated by the hypothesis that if a tag is useful for understanding user preferences (and thus for PE), the CF model will have learned a representation of it \citep{gopfert2021discovering}. After training a two-tower model, we use its item tower $\phi_I$ to learn CAVs. For a tag $g\in\calT$, we construct a training set $\bfD_g$ in which positive instances ($y=+1$) are items in $\bfT_g$ and negatives ($y=-1$) are those in $\bfT_{\olg}$.  
We then learn the CAV $\phi_g$ for $g$ by 
learning a (regularized) logistic regressor:
%
\begin{equation}
\label{eq:CAV-classifier}
\calL(\phi_g; \bfD_g) =  \sum_{(i,y)\in \bfD_g} \log(1 + e^{-y \phi_g^\top \phi_I(i)}) + \frac{\lambda}{2} \phi_g^\top \phi_g.
\end{equation}
%
The induced CAV $\phi_g$ is the normal to the separating hyperplane of this classifier, and offers a directional semantics for (the attribute corresponding to) tag $g$ in the item embedding space. Specifically, its dot product with $\phi_I(i)$, or \emph{$g$-score} $c_g(i)=\phi_g^\top\phi_I(i)$, quantifies the degree to which item $i$ satisfies tag $g$, (e.g., how violent or funny a movie is). We use the $g$-score to measure the {\em quality} (or usefulness) of CAV $\phi_g$, defining $Q(\phi_g;\bfD_g)$ to be the fraction of the item pairs $\{(i_1,i_2) \mid (i_1,+1)\in\bfD_g,(i_2,-1)\in\bfD_g\}$ for which $c_g(i_1)\geq c_g(i_2)$.
%
Given sparsity and noise in the tag data, we may be more confident in some CAVs than in others. We discuss 
CAV uncertainty below.

\vskip 1mm
\noindent
\textbf{2.4 \hskip 1mm Preference Elicitation.}
Traditional CF-based RSs uncover user preferences indirectly, using indirect feedback (e.g., clicks, ratings, purchases). \emph{Interactive RSs}, by contrast, actively elicit a user's preference by asking her questions, responses to which are used to refine its estimate (or \emph{belief}) about her utility function and improve quality of future recommendations \citep{he2016interactive}. At suitable points, an interactive RS selects a {\em query} $q$ from a {\em query space} $\calQ$ that reveals some aspects of the user's preferences. For example, in item-based PE, users may be asked to rate a particular item \citep{activecf:uai03,zhao13interactive}, in which case $\calQ=\calI$, or to choose an item from a slate $S$ that she most prefers \citep{viappiani:nips2010}, in which case $\calQ$ is the set of all subsets of $\calI$ of size $|S|$. In attribute-based PE \citep{chen_critiquing_survey:umuai2012}, a user can express her preferences with more abstract semantic attributes. 
We describe several types of elicitation queries
in Section~\ref{sec:queries-response-models}.

\vskip 1mm
\noindent
\textbf{2.5 \hskip 1mm Related Work.} 
Our work assumes an underlying CF method, and can be applied to matrix factorization methods \citep{salakhutdinov-mnih:nips07} or more recent DNN-based approaches \citep{yiEtAl:recsys19,yangEtAl:www20}. A number of approaches have been proposed for interpreting tags or attributes in RSs, often using tag or review data \citep{gantner:icdm2010,mcauley:icdm2012,luo2020latent}. Some works focused on learning tag semantics w.r.t.\ a learned recommendation model, including those that jointly learn both, or that derive tag meanings using a pre-trained CF model \citep{rendle_BPR:uai09,cohen:recsys2017,nema2021disentangling}. The model of \citet{gopfert2021discovering}, which we employ in this work, is of the latter form.

The literature on PE is substantial. Most related to our work are those which use Bayesian methods for eliciting user preferences~\cite{bourdacheEtAL:ijcai19,adam:uai21}. They maintain a \emph{belief state} over user preferences and refine it using PE queries. Item-based methods include, say, active CF methods that ask users to explicitly rate specific items \citep{activecf:uai03,zhao13interactive}, and item-selection methods that ask users to compare two items or state which item is most preferred within some set \citep{chajewska2000making,preference:aaai02,viappiani:nips2010}.
Attribute-based techniques include, for instance, example-critiquing schemes \citep{viappiani:06,chen_critiquing_survey:umuai2012}. Most work in PE deals with hard attributes, while we focus on soft 
attributes. One exception is the work of \citet{radlinski:sdd2019}, who develop a methodology for relating soft attribute usage in conversational RSs to user preferences.

\section{Queries and User Responses}
\label{sec:queries-response-models}

We first describe two novel \emph{query} types that we develop as well as the item queries. These queries can be used
to elicit information about a user's preferences for specific items or their (hard or soft) attributes. For each type, we develop \emph{response models} that relate a user's (stochastic) response to their underlying preferences, using the CAV semantics to interpret the user's behaviour w.r.t.\ attributes.
For ease of exposition, we assume that the CAV semantics is \emph{known} by both the RS and the user, but derive updates when there is uncertainty in the CAV semantics below. 
%


\noindent
\textbf{3.1 \hskip 1mm Attribute Queries.}
An \emph{attribute query} $q=(S,g)$ consists of a \emph{slate} of items $S$ and a tag $g$. The RS presents the slate to the user and asks her if she prefers items, relative to those in $S$, that are more/less---i.e., exhibit a greater degree of---$g$'s attribute (e.g., ``Would you prefer movies that are more/less thought-provoking than those in $S$?''). The user responds to positively $\rho=+1$ (i.e., more) or negatively $\rho=-1$ (i.e., less).  Slate $S$ can also be viewed as a recommendation set, not just a query (see Section~\ref{sec:active_search_of_queries}).

Our user response models for attribute queries assume user $u$ relates items in $S$ to her ``most preferred'' or \emph{target} item given her underlying utility function $\phi_u$.%
While $u$ would like to guide the RS to her most preferred item
$\argmax_{i\in\calI} \phi_u^\top \phi_I(i)$, this would assume an unrealistic level of familiarity with available items $\calI$ by $u$. Instead, we consider a model where $u$ targets a hypothetical \emph{ideal} item, $\phi^*_{I,u} \in \argmax_{\phi\in \Gamma} \, \phi_u^\top \phi$, 
w.r.t.\ some mildly constrained space $\Gamma\subset\calX$ unrelated to $\calI$.\footnote{For example, a user will not target a movie that is simultaneously maximally funny, though-provoking, and scary.} In lieu of a detailed familiarity model for $u$, we capture some familiarity with $\calI$ by constraining her \emph{target} item as follows:
 $\phi^*_{I,u} \in \argmax_{\phi\in \Gamma} \, \phi_u^\top \phi$,
 s.t.\ $\|\phi\|_2\leq \max_{i\in\mathcal I} \|\phi_I(i)\|_2$,
whose solution is
%
\begin{equation}
\label{eq:best_item_embed2}
\phi^*_{I,u} = \max_{i\in\mathcal I} \|\phi_I(i)\|_2 \cdot
\phi_u / \|\phi_u\|_2.
\end{equation}
%
The target item $\phi^*_{I,u}$ may fall outside of $\calI$, but this allows the user to direct the RS more meaningfully.

For attribute query $q=(S,g)$, in our \emph{mean-slate response model}, $u$ compares her target item $\phi^*_{I,u}$ with the mean embedding of items in $S$, $\phi_{I,\overline S}=\frac{1}{|S|}\sum_{i\in S} \phi_I(i)$. 
If the target exhibits more of attribute $g$ than the mean slate, i.e., the target's $g$-score $c_g(\phi^*_{I,u})$ is greater than $c_g(\phi_{I,\overline S})$, then
$u$ responds positively ($\rho=+1$). User response is estimated by the following \emph{probit} model \citep{train2009discrete,chaptini2005use}:
%
\begin{align}
P(\rho= +1 \mid q, \phi_u) 
&= \mathbf{\Phi}\left(c_g \left(\phi^*_{I,u} - \phi_{I,\overline S}\right)/\sigma_g\right),
\label{eq:BCAV_mean_slate}
\end{align}
%
where $\epsilon_g$ is a zero-mean Gaussian noise with variance $\sigma_g^2$,
and $\mathbf{\Phi}(\cdot)$ is the standard Gaussian CDF.

In our second \emph{mean-probability response model}, $u$ compares the $g$-score of her target with that of each item $i\in S$, and responds using the weighted average of differences:
\begin{align}
P(\rho = +1 \mid q,\phi_u) 
&= \frac{1}{|S|} \sum_{i\in S} \lambda_i \cdot \mathbf{\Phi}\left(c_g \left(\phi^*_{I,u} - \phi_{I}(i)\right)/\sigma_g\right)\;,
\label{eq:BCAV_mean_prob}
\end{align}
where $\lambda_i>0$ is a response weight
s.t.\ $\sum_{i\in S}\lambda_i=1$.
While mean-slate model is intuitive, the flexibility of mean-probability may better capture nuances in user behavior (see \emph{IpA} below).

\noindent
\textbf{3.2 \hskip 1mm Item Queries.}
An \emph{item query} (or choice query) $q=S$ presents a slate $S$ to $u$ and asks her which item is preferred \citep{ben1985discrete,viappiani:nips2010}. User response $\rho=i$
is given by a standard \emph{multinomial logit} model w.r.t.\ $\phi_u$ \citep{chaptini2005use,ben1985discrete}: 
\begin{equation}
P(\rho = i \mid q, \phi_u) = \frac{\exp(\phi_I(i)^\top\phi_u/T)}{\sum_{j\in S} \exp(\phi_I(j)^\top\phi_u/T)},
\end{equation}
%
where $T$ is a temperature parameter.
%

\noindent
\textbf{3.3 \hskip 1mm Item-plus-Attribute Queries.}
\emph{Item-plus-Attribute (IpA) queries} combine attribute and item queries, and have the same form $q=(S,g)$ as attribute queries. The user
is first asked to select her preferred item $i^\ast_S$ in $S$ (per item queries), then asked to critique $i^\ast_S$ w.r.t.\ $g$
(per attribute queries).
The user responds with $\rho=(i,+1)$ or $\rho=(i,-1)$ by comparing 
$\phi^*_{I,u}$ to $i^\ast_S$ (rather than the slate) via $g$-scores, and otherwise responds as in attribute queries:
%
\begin{align}
\label{eq:response_pref_CAV_query}
&P(\rho_1 = i \mid S,\phi_u) \!\times\! P(\rho_2 = y \mid g,\phi_u, \rho_1=i) \nonumber \\ 
&\!=\!\frac{\exp(\phi_I(i)^\top\phi_u/T)}{\sum_{j} \exp(\phi_I(j)^\top\phi_u/T)} \!\times\! \mathbf{\Phi}\!\left(\!y \cdot c_g \phi^*_{I,u} \!-\! \phi_{I}(i) / \sigma_g\!\right). 
\end{align}

\noindent
\textbf{3.4 \hskip 1mm CAV Uncertainty.} 
In contrast to hard attributes, the semantics of soft attributes are typically estimated using sparse, noisy data (e.g., as in our use of tag data above). As a result, CAVs constructed for different tags may be characterized by varying degrees of uncertainty. For instance, if CAV $\phi_g$ for tag $g$ is based on a large amount of tag data with little disagreement in usage, the RS should be more confident in $\phi_g$ than in the CAV $\phi_{g'}$ for a tag $g'$ whose data is sparser or less consistent. Abstractly, we assume that the RS has 
a 
\emph{CAV belief (distribution)} $P_g(\phi_g \mid \bfD_g)$ reflecting this uncertainty (where $\bfD_g$ is the tag data used to train $\phi_g$). We do not require a specific mechanism for generating this belief,
but Bayesian logistic regression (augmenting Eq.~\ref{eq:CAV-classifier}) \citep{jaakkola1997variational} or Bayesian learning-to-rank \citep{kuo2009learning} are suitable methods.

The response models for attribute and IpA queries above depend on the CAV/semantics of the attribute/tag in question. In what follows, we assume that user $u$ responds using some true, underlying CAV $\phi_g$.\footnote{Different users may have 
different interpretations of an attribute. Such \emph{subjectivity} can be uncovered using CAVs \citep{gopfert2021discovering}, but we do not consider this here.} To reflect this additional uncertainty in a user response, the RS interprets it w.r.t.~its CAV belief. Specifically, $P(\rho\mid q, \phi_u)$ in Eqs.~\ref{eq:BCAV_mean_slate} and~\ref{eq:response_pref_CAV_query} requires taking an expectation over possible CAVs $\phi_g$ w.r.t.\ $P_g(\phi_g | \bfD_g)$. These response probabilities are then used to update RS beliefs about $u$'s utility (Section~\ref{sec:learning_from_queries}) and to choose queries (Section~\ref{sec:active_search_of_queries}) when the CAVs are uncertain.

Specifically, $P(\rho\mid q, \phi_u)$ in Eqs.~\ref{eq:BCAV_mean_slate}, \ref{eq:BCAV_mean_prob} and~\ref{eq:response_pref_CAV_query} requires taking an expectation over possible CAVs $\phi_g$ (or equivalently, scoring functions $c_g$) w.r.t.\ $P_g(\phi_g | \bfD_g)$. These response probabilities are then used to update beliefs about $u$'s utility (Section~\ref{sec:learning_from_queries}, by Bayesian methods) and selecting optimal queries (Section~\ref{sec:active_search_of_queries}) when (some or all) the CAVs are uncertain.
See Appendix~\ref{app:CAVuncertainty} for further details.



\section{User Belief State Modeling}
\label{sec:learning_from_queries}

Bayesian methods 
generally maintain a \emph{belief state} or distribution  over a user’s utility function
\cite{vendrov2020gradient,preference:aaai02}. 
In our setting, the belief state for user $u$ is initially $P_U(u)$, and is refined as $u$ responds to PE queries.
Given response $\rho$ to query $q$, we update our belief about $u$ in the standard Bayesian fashion:
$$P_U(u \mid \rho,q) := P(\phi_u \mid \rho,q) \propto P(\rho \mid q,\phi_u) P_U(u).$$ 
The prior $P_U(u)$ is the Gaussian user embedding learned by our two-tower model (Section~\ref{sec:formulation}). Let $\calH^{(K)} = \{(q^{(K)},\rho^{(K)})\} \cup\calH^{(K-1)}$ with $\calH^{(0)} = \emptyset$, be a history 
of $K$ queries and user responses. Given $\calH^{(K)}$, the RS's posterior for $u$ is 
\begin{equation}
\label{eq:posterior_belief}
P_U(u\! \mid\! \calH^{(K)})\! =\! P(\phi_u\! \mid\! \calH^{(K)})\! \propto\! P_U(u) \prod_{k=1}^{K} P(\rho^{(k)}\! \mid\! q^{(k)},\!\!\phi_u).
\end{equation}
This assumes conditional independence of 
responses
given $\phi_u$.
Unlike the prior, generally the posterior is not Gaussian, and thus, we use the following two methods to tractably approximate it.


\subsection{Parameterized Posterior}
\label{subsec:param_posterior}
We can approximate the posterior with some parameterized distribution $P(u;\theta)$. 
To do so, we sample from the true unnormalized posterior using Metropolis-Hastings~\cite{chib1995understanding} or Hamiltonian Monte Carlo (HMC). These samples $\{\phi_{u,i}\}_{i=1}^n$ are then used to estimate parameter $\theta$ 
by maximizing the log-likelihood
\begin{align}
L(\theta \mid \calH^{(K)})&=\sum_{i=1}^n \log P(\phi_{u,i} \mid \calH^{(K)};\theta) \label{eq:parameterized1} \\ 
&=\sum_{i=1}^n \log P(\rho^{(K)}|q^{(K)}\! , \phi_{u,i};\theta)\! +\! \log P(\phi_{u,i}|\calH^{(K-1)}\! ;\! \theta). \nonumber
\end{align}
We test two different variants of the sampling and posterior update methods in this scheme. In the first \emph{batch} method, we generate a large set of samples and use them to update the posterior offline, i.e.,~for each query no additional sample of posterior belief is generated before the posterior is updated. In the second \emph{iterative} method, we generate fewer samples from our initial posterior, update the posterior with these samples, and then re-generate new samples with this updated posterior. This process is repeated several times for each query. While the iterative method is more computationally expensive, our experiments (Section~\ref{sec:simulations}, Figure~\ref{fig:posterior_update}) demonstrate its improved data efficiency over the batched counterpart.

\subsection{Gaussian Posterior}
The posterior can be ``assumed'' to be Gaussian $\calN(\hat{\phi}_{\mu,U}(u),\hat{\phi}_{\sigma,U}(u))$ using the Laplace approximation~\cite{williams2006gaussian,biyik2020active,li2021roial}. For this, we treat the posterior mean $\hat{\phi}_{\mu,U}(u)$ as the mode of the true log-posterior, i.e.,
\begin{align}
\hat{\phi}_{\mu,U}(u) &\in \argmax_{\phi\in\mathbb R^d} \Big[\sum_{k=1}^K \log P\big(\rho^{(k)} \mid q^{(k)},\phi\big) \label{eq:laplace1} \\
&-\frac12 \big(\phi - \phi_{\mu, U}(u)\big)^\top \big(\phi_{\sigma, U}(u)^\top \phi_{\sigma, U}(u)\big)^{-1}\big(\phi - \phi_{\mu, U}(u)\big)\Big]. \nonumber
\end{align}
This optimization may not be convex, but its local optima can be found efficiently because its gradient can be written in closed-form. 
$\hat{\phi}_{\mu,U}(u)$. In Figure~\ref{fig:attr_categorical_gaussian_cosine} in Section 6, we numerically compare the performance of the Gaussian posterior with the parameterized (categorical) posterior. The discrete posterior model outperforms its Gaussian counterpart with both attribute and IpA response models. We therefore use the parameterized posterior as the default belief state model for all PE methods described below.

\subsection{CAV Uncertainty}
When the CAV used by an attribute or IpA query is uncertain, the response probabilities used in our belief state updates (Eqs.~\ref{eq:posterior_belief}, \ref{eq:laplace1} or~\ref{eq:parameterized1}) are computed using expectations over CAV samples w.r.t. its belief distribution $P_g(\phi_g | \bfD_g)$ (see Section 3.4). Similarly, we update the belief about $u$ with responses:
$$P_U(u \mid \rho,\tilde{q}) := P(\phi_u \mid \rho,\tilde{q}) \propto P(\rho \mid \tilde{q},\phi_u) P_U(u).$$ 
Given any query-response pair $(\tilde{q},\rho)$ under the current belief state $P_U(u)$, the posterior belief is updated by Bayes rule:
\begin{equation}
    P(\rho \mid \tilde{q}, \phi_u) = \frac{P(\rho \mid \tilde{q},\phi_u) P(\phi_u)}{\int_{\phi_u} P(\rho \mid \tilde{q},\phi_u) P(\phi_u) d\phi_u}.
\end{equation}



\section{Query Optimization}
\label{sec:active_search_of_queries}

In this section, we develop methods to optimize the choice of queries. 
A key objective is to find queries whose responses quickly refine the posterior 
to improve recommendation quality. However, the RS may also want to use queries whose slates contain ``good'' recommendations w.r.t.\ the current posterior, not just ``good'' information for posterior update, since the user may select/consume one of these items at any time. We address the trade-off between information gathering and recommendation quality below.\footnote{For item queries without attributes, the optimal recommendation and query slates
are identical under several natural response models, resolving this tension \citep{viappiani:nips2010}.}




\subsection{Pure Preference Elicitation}

In \emph{pure preference elicitation (PPE)}, we focus on the \emph{information} a query response provides about user $u$'s utility (and ignore the predicted utility of items in $S$). 
We consider several \emph{acquisition functions (AFs)} to find the query with the most useful information w.r.t.\ reducing uncertainty in
the belief state $P_U(u)$.%
\footnote{An ideal AF would be a \emph{policy} that optimizes the \emph{sequence} of queries \citep{preference:aaai02,white:ejor03}. However, such a sequential formulation is generally intractable; thus, we consider \emph{myopic} approaches and optimize the AF w.r.t.\ the immediate query only (as is common in PE \citep{chajewska2000making}).} 


\noindent
\textbf{Random Query Selection:} 
This is a natural baseline that samples queries uniformly at random
from $\calQ$.

\noindent
\textbf{Entropy:}
To make posterior beliefs more informative, 
one can use conditional Shannon entropy~\citep{cover1999elements}: 
 \begin{equation}
      H(\rho \mid q,\calH) = -\mathbb E_{\rho | q,P_U(u | \calH)}[\log P(\rho | q,P_U(u | \calH))].
 \end{equation}
Computing the expected posterior entropy for a given query in closed-form is generally hard, but one can estimate it by taking an expectation over all sampled posterior user responses. This metric measures the amount of latent user information that
remains after a PE query $q$. 

\noindent
\textbf{Mutual Information:}
We can measure the information content of $q$ using \emph{mutual information (MI)} between $u$'s target $\phi_{I,u}^*$ and her response $\rho$, given $q$ and $\calH$:  
\begin{equation}
\label{eq:MI}
MI(\phi_{I,u}^* ; \rho \!\mid\! q,\calH) \!:=\! H(\rho \!\mid\! q,\calH) \!-\! \mathbb{E}_{\phi_{I,u}^*\!\mid\! \mathcal H}\big[H(\rho \!\mid\! q,\phi_{I,u}^*)\big].
\end{equation}
We can also express the distribution over target items as $P(\phi_{I,u}^* | \calH) = \int P(\phi_{I,u}^*,\phi_u | \calH)d\phi_u = \mathbb{E}_{\phi_u\sim P_U} \!\big[P(\phi^*_{I,u} | \phi_u)\big]$.
\\Using \eqref{eq:best_item_embed2}, the second term is equal to $\mathbb{E}_{\phi_{I,u}^* | \calH}\big[H(\rho | q,\phi_{I,u}^*)\big] =  \mathbb{E}_{\phi_u\sim P_U(u|\calH)}\big[H(\rho | q,\max_{i\in\calI}\frac{\|\phi_I(i)\|_2 \cdot\phi_u}{\|\phi_u\|_2} )\big]$,
%
which is estimated by sampling $\phi_u\sim P_U(u \mid \calH)$.
We choose a query that minimizes this measure to 
make our posterior model correlated with $u$'s target.

\noindent
\textbf{Expected Value of Information (EVOI).}
The EVOI acquisition function \citep{chajewska2000making,preference:aaai02,viappiani:nips2010} measures the improvement in $u$'s expected utility resulting from a response to query $q$. The EVOI of $q$ (given $\calH$) is
\begin{equation}
\label{eq:evoi_definition}
\EVOI(q \mid \calH) = \PEU(q \mid \calH) - \EU^*\big(P_U(u \mid \calH)\big), 
\end{equation}
where 
\begin{equation}
\label{eq:eu_definition}
EU^*(P_U(u)) = \max_{i\in \calI} \; \mathbb{E}_{\phi_u \sim P_U(u)}\big[\phi_u^\top\phi_I(i)\big]
\end{equation}
and $PEU(q \mid \calH)$ is the \emph{posterior expected utility} of $q$, 
\begin{equation}
\label{eq:peu_definition}
PEU(q \mid \calH) = \sum_{\rho} P(\rho \mid q,\calH) \cdot EU^*\big(P_U(u \mid \calH\cup\{(\rho,q)\})\big).
\end{equation}
%
EVOI extracts information that offers maximum expected improvement in recommendation quality. It can be also used to decide when to stop elicitation (e.g., once it falls below some tolerance or exceeds the cost of interaction).
The direct computation of PEU can be expensive, so we approximate it by sampling from the prior $P_U(u\mid\calH)$. With CAV uncertainty, the response probabilities used in PEU and other AFs are computed using expectation over CAV. The details are deferred to App.~\ref{app:CAVuncertainty}.

\noindent
\textbf{CAV Uncertainty.}
With CAV uncertainty, response probabilities used in PEU and other AFs are computed using expectation over CAVs as discussed above. For example,  the EVOI of a soft attribute query $\tilde{q}$ (given $\calH$) is:
\begin{equation}
\EVOI(\tilde{q} \mid \calH) = \PEU(\tilde{q} \mid \calH) - \EU^*\big(P_U(u \mid \calH)\big).
\end{equation}
where $EU^*(P_U(u))$ is still the same as Eq (10)
but $PEU(\tilde{q} \mid \calH)$ is the \emph{posterior expected utility} of w.r.t.\ $\tilde{q}$,
\begin{equation}
PEU(\tilde{q} \mid \calH) = \sum_{\rho} P(\rho \mid \tilde{q},\calH) \cdot EU^*\big(P_U(u \mid \calH\cup\{(\rho,\tilde{q})\})\big).
\end{equation}
Other acquisition functions can also be extended analogously but their details will be omitted for the sake of brevity. 


\subsection{Blended Elicitation \& Recommendation}

In \emph{blended PE and recommendation (BPER)}, we include the quality of the slate $S$ used in query $q$
when assessing $q$, not just its information value.
BPER blends two objectives, {\em information gathering} $\IG(q | \calH)$ and {\em recommendation quality} $\RQ(q | \calH)$. $\IG(q | \calH)$ measures the utility information extracted by $q$ and can use any pure PE AF (e.g., Entropy, MI, EVOI).
$\RQ(q | \calH)$ measures slate quality (i.e., expected utility) of $S$: $\RQ(q |\calH):=\sum_{i\in S}\mathbb{E}_{\phi_u\sim P_U(u\ | \calH)}[\phi_u^\top\phi_I(i)]$. 
To balance
the two, 
we define the BPER AF as $\gamma \IG(q | \calH) + (1 - \gamma)\RQ(q | \calH)$, where $\gamma\in[0,1]$. 
We expect an RS to focus on IG earlier in an interaction sequence to learn about $u$, and gradually shift focus to RQ. While decaying $\gamma$ seems natural, experiments suggest that a well-tuned \emph{constant} $\gamma$ suffices, since IG 
tends to decrease over time as the posterior
converges to $u$'s true embedding, after which the BPER AF would focus more on recommendation. 

\subsection{Query Optimizers}

Query selection requires optimizing the chosen AF
over query space $\calQ$. The size of $\calQ$ depends linearly on the number of tags $|\calT|$ and combinatorially on the number of items $|\calI|$, where usually $|\calT| << |\calI|$. Optimal tags can be found by searching over $\calT$, while slate optimization is demanding when $\calI$ is large, hence requires approximation.
We consider two approaches to slate-attribute selection: optimizing slate $S$ (or an item inside $S$) and then tag $g$; and optimizing both jointly. 



\noindent
\textbf{Thompson Sampling.}
Slate $S$ is constructed using sequential TS \citep{gopalan2014thompson}: $S_0 = \emptyset$ and $S_j = S_{j-1} \cup \argmax_{i\in\calI\setminus S_{j-1}} \phi_u^\top\phi_I(i)$, where $\phi_u \sim P_U(u\mid\calH)$ and $j\in\{1,\ldots,\lvert S\rvert\}$.
Given $S$, $g$ is randomly selected.

\noindent
\textbf{Sequential Greedy.}
Each item in the slate is selected greedily over the item set (excluding items that have already been added).
The first item in $S$ is the ``best'' item; at each subsequent step, we alternate between updating $g$ and the next item in $S$ using the AF until $S$ is complete.

\noindent
\textbf{Random Search.}
We randomly generate a certain number of slates-attribute pairs, then apply the AF to select the query with the highest score. With CAV uncertainty, we compute score of each query by averaging over sampled CAVs.

\noindent
\textbf{Continuous Relaxation.}
Inspired by gradient-based PE in~\cite{vendrov2020gradient}, we relax the combinatorial query selection problem into a continuous one and solve it using first- or second-order methods. Given this continuous representation of $q$, we project it back onto the true query space.
To recover $S$ and $g$, we project using Euclidean distance. The main challenge in continuous optimization lies in the discontinuous nature of our AFs, which require maximizing over the discrete item set. We circumvent this with a normalization assumption
(see Appendix~\ref{app:gradient_of_acquisition_wrt_slate}).
With CAV uncertainty, since tag $g$ is a random variable, we have to represent $P_g$ by a multivariate normal distribution and apply the re-parameterization trick (see Appendix~\ref{app:relaxation}). We project $P_g$ using KL-divergence.

\section{Empirical Results}
\label{sec:simulations}

We conduct experiments to test the effectiveness of our PE methods, combining various query types, belief update methods, query selection schemes, and slate optimization methods. We evaluate our algorithms in three domains: (i) a simple synthetic environment; (ii) a complex simulated environment similar to that
used by \citet{gopfert2021discovering}; and (iii) one derived from the MovieLens 20M dataset \cite{harper16:movielens}. The simulated domains provide us with ground-truth user utility and attribute semantics to allow precise evaluation. We describe the environments, our evaluation metrics, then a set of experimental studies in each of the three domains.

\noindent
\textbf{Synthetic Environment.}
We set $\lvert\calI\rvert=1000$, $\lvert\calT\rvert=10$, with item embeddings $\phi_I(\cdot)$ and CAV vectors $\phi_g$ 
sampled from a $d$-dimensional ($d=5$) Gaussian $\mathcal N({\bf 0},{\bf I})$. Each user is represented by a Gaussian distribution $P_U$ with a random mean vector and covariance matrix. User response noise is $\sigma_g = 0.1$ for all $g\in\calT$.

 

\noindent
\textbf{RecSim NG Environment.}
Following \citet{gopfert2021discovering}, we use the RecSim NG \cite{mladenov2020recsimng} environment to construct a user model to generate ratings and tags, and then use the resulting dataset for learning CAVs and for PE. With $\lvert\calU\rvert=25,\!000$ users and $\lvert\calI\rvert=10,\!000$ items, we represent each user and item in a $d$-dimensional embedding space ($d=25$)---each item dimension reflects a latent ``attribute'' and each user dimension a utility for that item attribute. User ratings for items are generated by a staged sampling process which approximates the user/item-embedding dot product (plus noise and rating discretization). 
The number of ratings given by users follows a power law distribution to ensure ratings sparsity.
Five of the $25$ latent attributes are taggable ($\lvert\calT\rvert=5$). Users can only tag rated items and are more likely to tag higher-rated items. For any  tag $g$, a \emph{fixed} threshold $\tau_g$ gives the probability of $u$ tagging the item; we set $\tau_g=0.5$ for soft attributes. We train CAVs for each tag with logistic regression using the item-tag data. Average CAV quality (accuracy) is $0.909$ on the test set, and Spearman correlation between predicted and ground-truth tags is $0.570$. Noise is $\sigma_g = 0.25$ for all $g\in\calT$. The temperature $T$ in the item response model is $0.5$.


\noindent
\textbf{MovieLens 20M.}
We also evaluate our methods on the more realistic MovieLens 20M dataset \cite{harper16:movielens}. There are $465$K tag-instances in which $138K$ users applied tags to $27K$ movies. Tags mainly represent movie genres (e.g., action, drama) or more subjective descriptions (e.g., quirky, funny). Following the methodology of \cite{gopfert2021discovering},
we split ratings and tag data into train and test sets such that any user-item pair is present exactly in one of these sets. We generate $d$-dimensional ($d=50$) user and item embeddings using alternating least-squares (ALS) and train CAVs on this latent space. 
Due to item-tag sparsity, we train CAVs only for the $164$ most-frequently used tags (w.r.t.\ unique users, items).
Average CAV test quality is $0.727$. User response noise is again $\sigma_g = 0.25$ and the temperature in the item response model is $T = 0.5$.

\noindent
\textbf{Metrics.}
We use three metrics to evaluate the performance of our PE methods. (i) \textbf{Cosine} is the cosine similarity (or alignment \cite{sadigh2017active}) between the mean user posterior embedding and true user embedding. After $k$ queries, it is given by $\textbf{Cosine} = \frac{\phi_u^\top \; \mathbb{E}_{\phi\sim P_U(\cdot\mid \calH^{(k)})}[\phi]}{\|\phi_u\|_2\cdot \|\mathbb{E}_{\phi\sim P_U(\cdot \mid \calH^{(k)})}[\phi]\|_2}\:$.
(ii) \textbf{NDCG} is the normalized discounted cumulative gain \cite{jarvelin2002cumulated} between the true top $\lvert S\rvert$ items and the top $\lvert S\rvert$ items estimated using the posterior. (iii) \textbf{Query NDCG} is a variant of NDCG where the slate $S^{(k)}$ that is \emph{presented} to the user in the $k$'th query replaces the \emph{a posteriori} optimum. We use this measure primarily with the BPER scheme to study the trade off between IG and RQ. 

\noindent
\textbf{Implementation Details.}
We implement our PE algorithms using Tensorflow~\cite{tensorflow} which offers automatic differentiation for both Laplace approximation and continuous relaxation. We use Tensorflow Probability~\cite{dillon:tfp} for HMC in parameterized posterior updates and for probability/likelihood computation.

\begin{figure}
    \centering
    \includegraphics[width=1.0\textwidth]{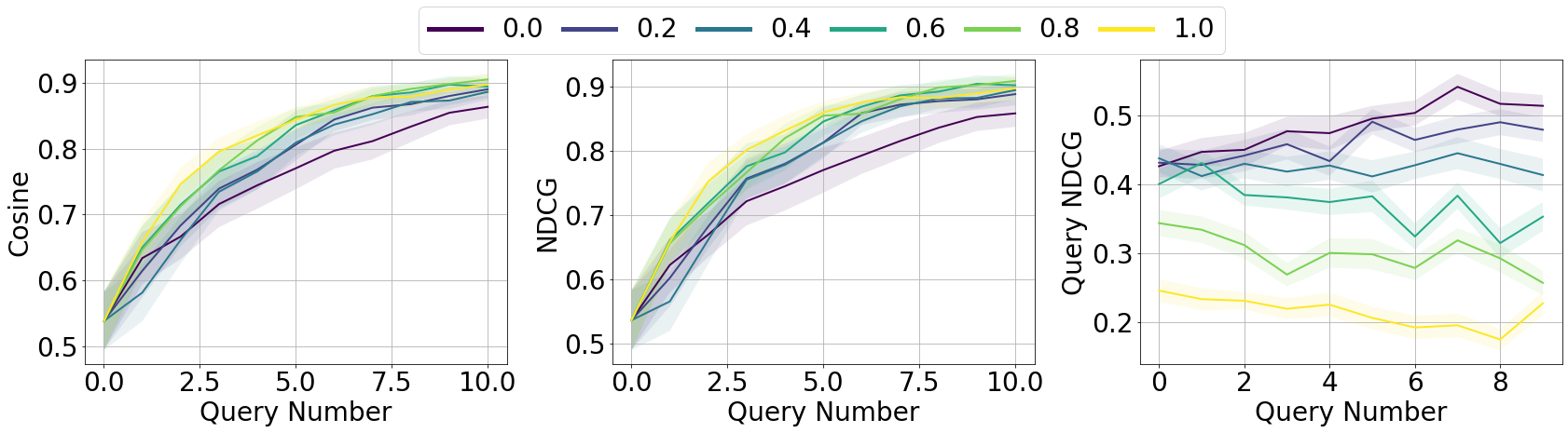}

    \caption{IG, RQ ($\gamma$) tradeoff, BPER scheme (Synthetic).}

    \label{fig:gammas}
\end{figure}

\noindent
\textbf{Experiment 1: IG-RQ Tradeoff with Synthetic Data.} 
We use the synthetic environment to assess the IG-RQ tradeoff and its effect on PE performance. We initialize the RS with the prior user embedding distribution, after which it queries a simulated user 10 times with a slate of size 5 (and tag if needed). We test each PE method with $10$ users, $5$ runs each (different random seeds). We report the mean and standard deviation of our metrics over these $50$ runs.
Figure~\ref{fig:gammas} shows the trade off between IG and RQ in the BPER scheme. Using IpA queries, EVOI as our AF, and random-search slate optimization, we run PE with $\gamma$ ranging from $1$ (pure elicitation) to $0$ (pure recommendation). While tuning $\gamma$ offers only a modest improvement in elicitation quality (see cosine and NDCG), PE with smaller $\gamma$ generally places higher quality items on the slate during elicitation, with quality improving with the number of queries (see Query NDCG). The nonlinear trade-off between IG and RQ induces a  ``sweet spot'' at around $\gamma=0.5$, and therefore we use this $\gamma$ for all subsequent experiments. 



\begin{figure}[H]
    \centering
\includegraphics[width=0.5\textwidth]{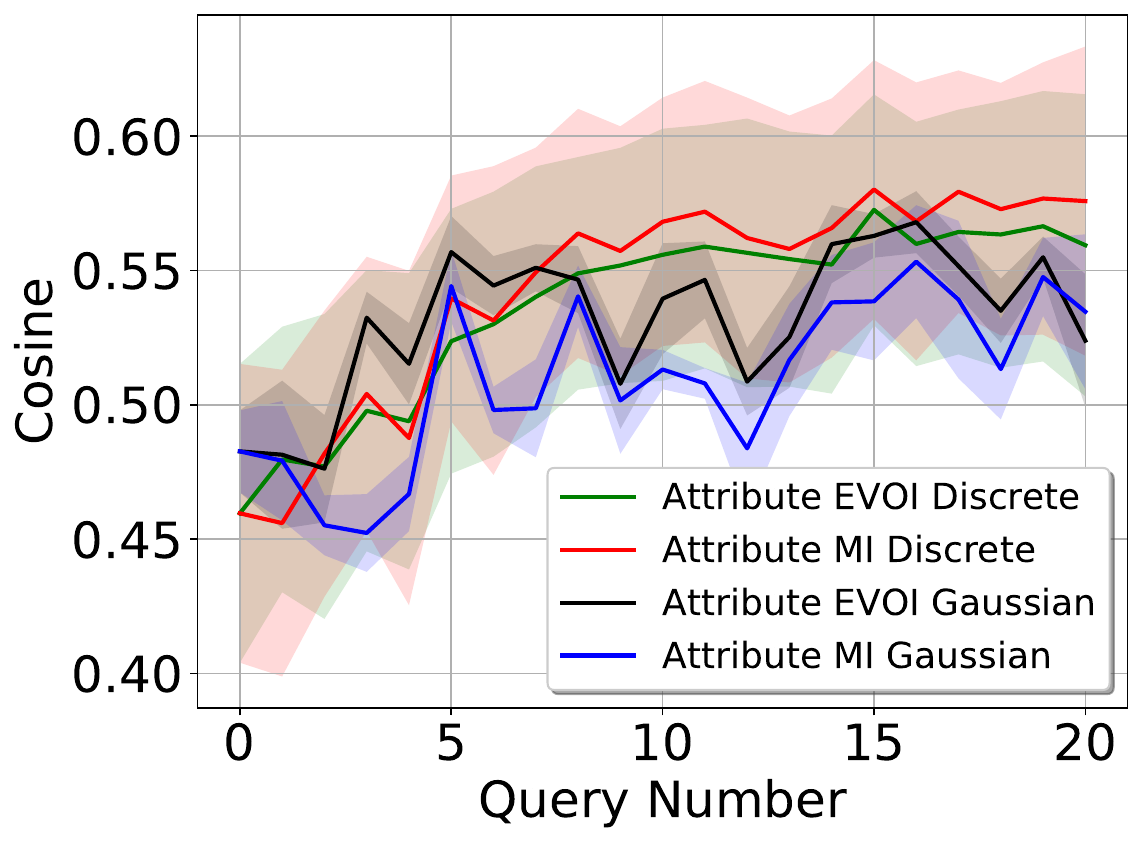}\,\includegraphics[width=0.5\textwidth]{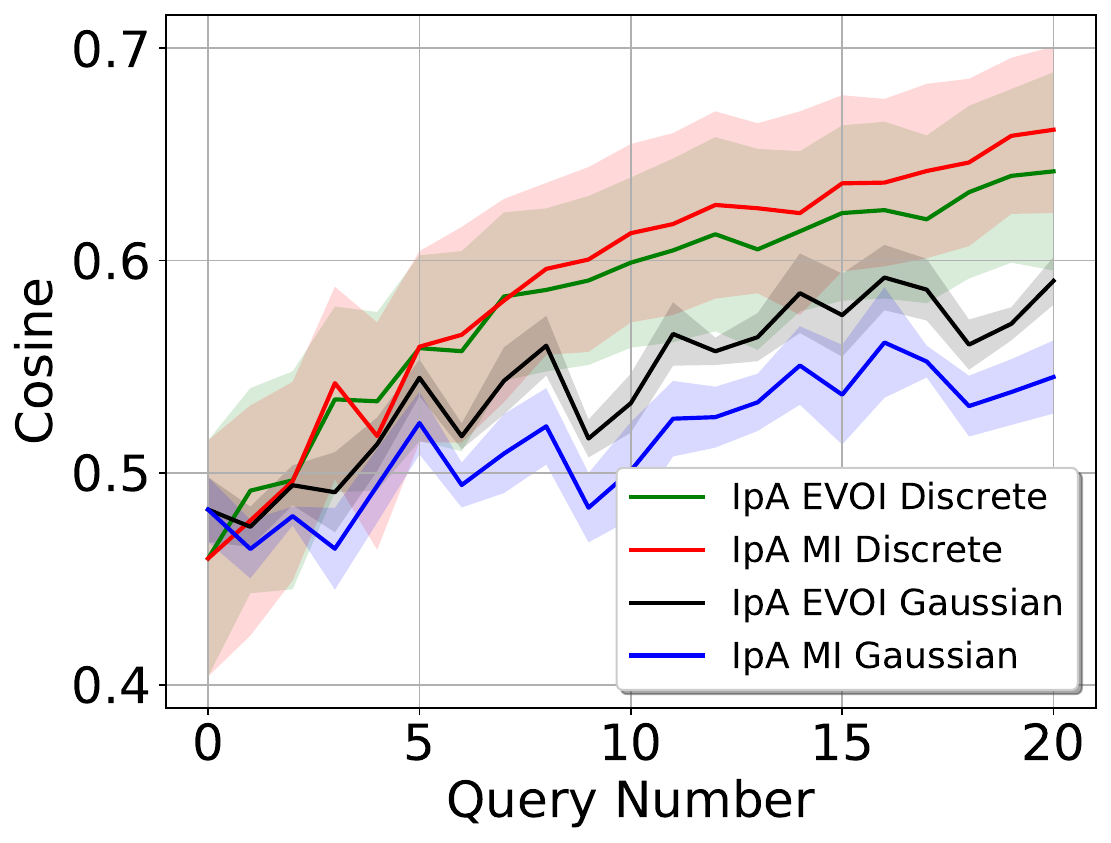}

\caption{Cosine similarity on RecSim NG with Gaussian (left) and parameterized (right) posterior.}

\label{fig:attr_categorical_gaussian_cosine}
\end{figure}

\noindent
\textbf{Experiment 2: Posterior Update Methods with RecSim NG.} In this experiment we compare the performance of the Gaussian posterior, updated via the closed-form Laplace approximation, with the parameterized (categorical) posterior, updated using HMC as specified in Section 4.2. 
Figure~\ref{fig:attr_categorical_gaussian_cosine} presents the performance of different posterior update methods in attribute-based PE, tested in the Recsim NG environment. 
We compare the accuracy of belief states via cosine similarity between the current posterior embedding and the underlying ground-truth one. Clearly, parameterized posterior models outperform the Gaussian counterpart in both attribute and IpA response models.

\begin{figure}
    \centering
    \includegraphics[width=0.5\textwidth]{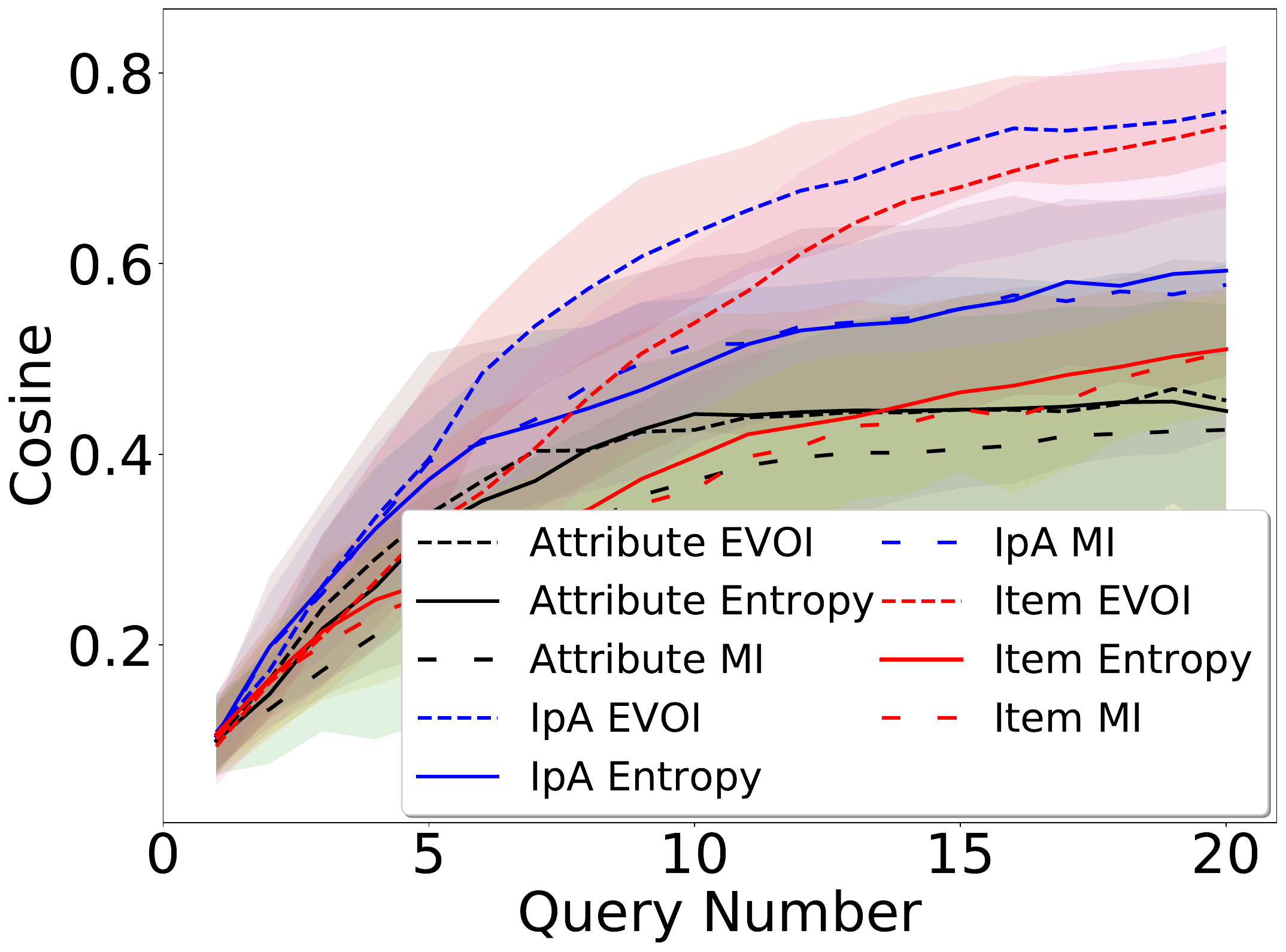}\,\includegraphics[width=0.5\textwidth]{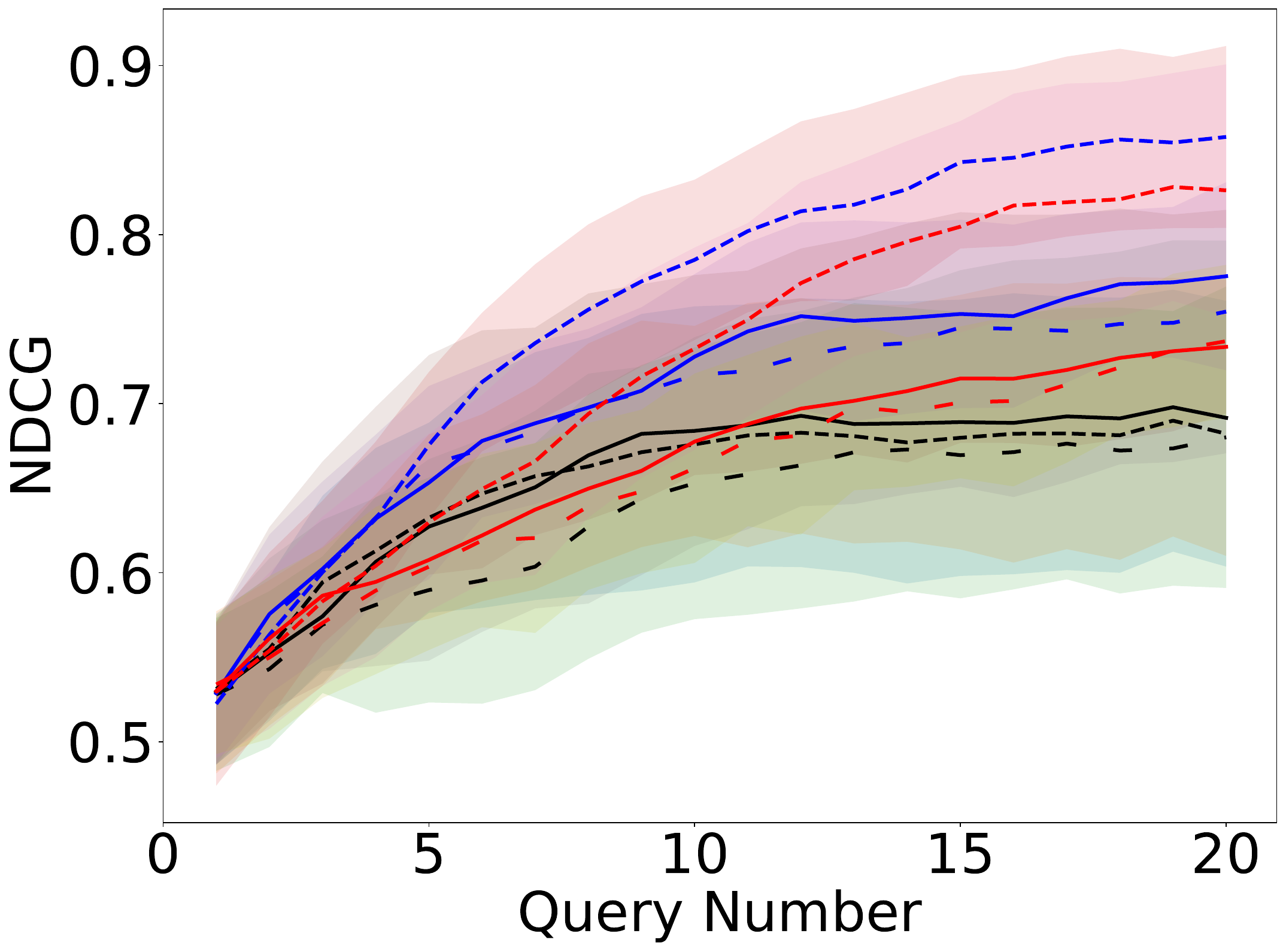}

    \caption{Comparing query types and AFs (RecSim NG).}

    \label{fig:acquisitions}
\end{figure}

\noindent
\textbf{Experiment 3: Ablation Studies with RecSim NG.}
Using the RecSim NG environment, we set the RS prior over user embeddings to be Gaussian (Section~\ref{sec:formulation}). The RS queries each user $20$ times using slates of size $5$ (for each query type).
We test each PE method with $16$ random users over $5$ runs. We run several ablation studies to assess how different user query responses, query/slate optimizers, and AFs affect information gathering (IG) and recommendation quality (RQ). 
We focus on the cosine and NDCG metrics (as Query NDCG mainly measures the IG-RQ trade-off). To focus on the CAV/query-type interaction, our first four studies assume no CAV uncertainty. Figure~\ref{fig:acquisitions} compares the different query types and AFs, while fixing the query optimization method to be random search over $100$ queries. Among the query types, PE with IpA performs the best, followed by item then attribute queries. This is unsurprising since IpA elicits the most information. While attribute queries provide the simplest user feedback, PE with attribute queries is quite effective initially, achieving similar performance to item and IpA-based PE; though with only $5$ taggable attributes, performance quickly saturates with more queries. EVOI is the most effective AF, outperforming MI and Entropy, and generating higher quality recommendations at each stage of the interaction (number of queries).
This reflects the fact that EVOI-maximizing queries elicit user utility information directly aimed at improving recommendation quality.
 
\begin{figure}
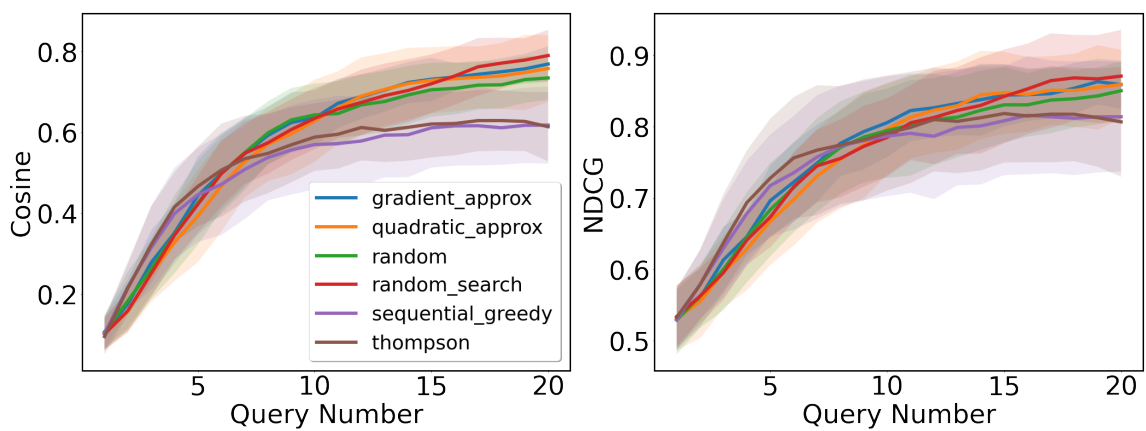

    \centering
\includegraphics[width=0.5\textwidth]{figures/synthetic_yinlam_acquisition_method_cosine.pdf}\,\includegraphics[width=0.5\textwidth]{figures/synthetic_yinlam_acquisition_method_ndcg.pdf}

    \caption{Comparing query optimizers (RecSim NG).}

    \label{fig:slates}
\end{figure}

In Figure~\ref{fig:slates}, we assess the impact of various query optimizers on PE and recommendation quality. We use IpA queries, the EVOI AF, and BPER with $\gamma=0.5$. We consider five query optimizers: (i) random search; (ii) continuous relaxation with first-order optimization; (iii) continuous relaxation with second-order optimization; (iv) sequential greedy (myopic w.r.t.\ BPER AF); and (v) Thompson sampling (TS). We also use a baseline in which both the slate and tag are selected uniformly at random. Random requires the least computation, followed by sequential greedy, TS, and then the joint optimizers on slates and tags. The additional computation cost of random search and continuous relaxation offers significant gains in IG and RQ, with a 10--15\% NDCG improvement over TS and sequential greedy (though TS and sequential greedy perform well early in the PE process). While joint-optimization-based methods still outperform random, random fares better than the greedy methods, TS, and sequential greedy, which is surprising given its simplicity. We conjecture this is because the RecSim NG environment is quite simple, and the inherent item diversity on random slates itself helps reduce belief state entropy.


\begin{figure}
    \centering
\includegraphics[width=0.5\textwidth]{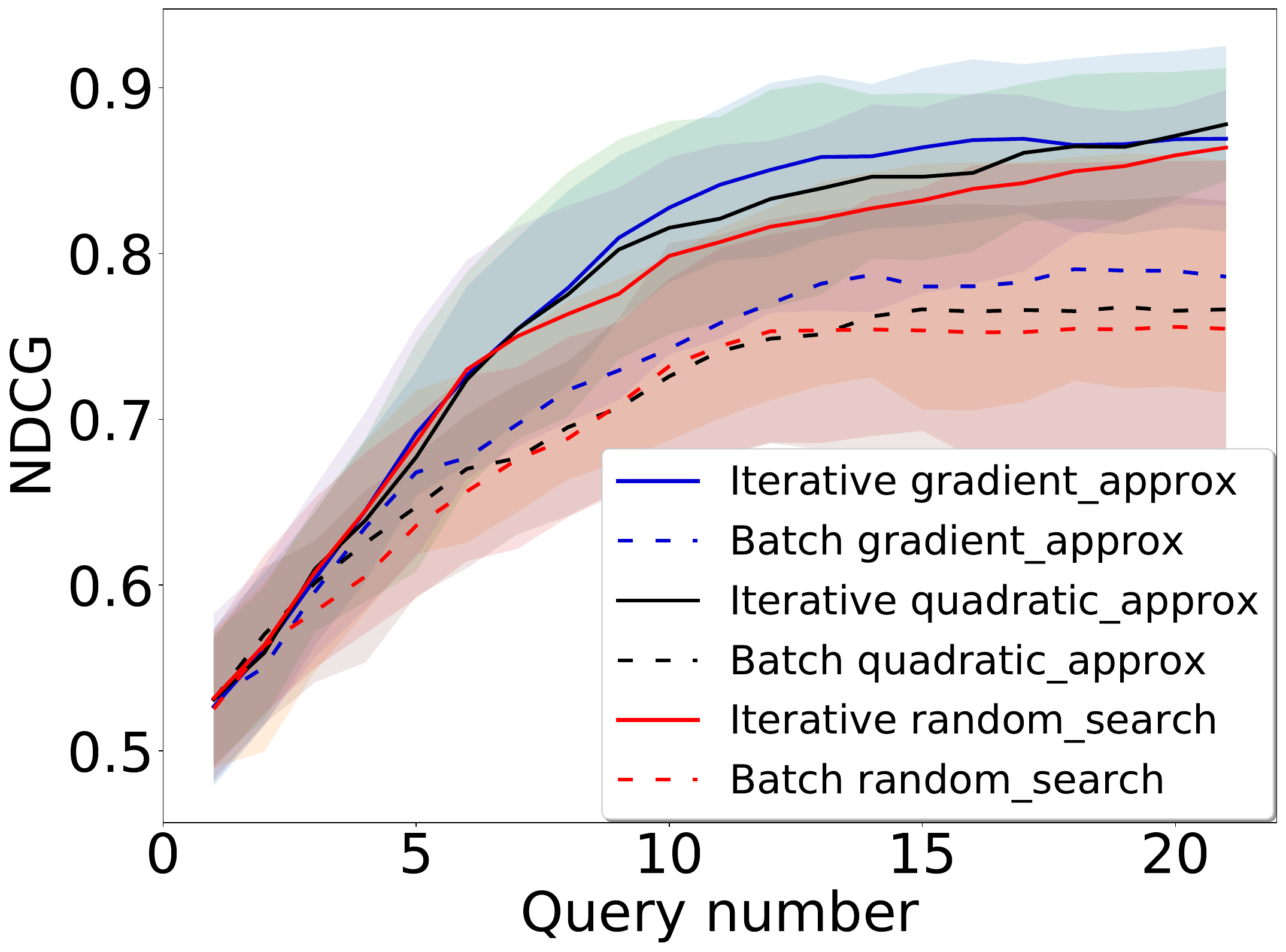}\,\includegraphics[width=0.5\textwidth]{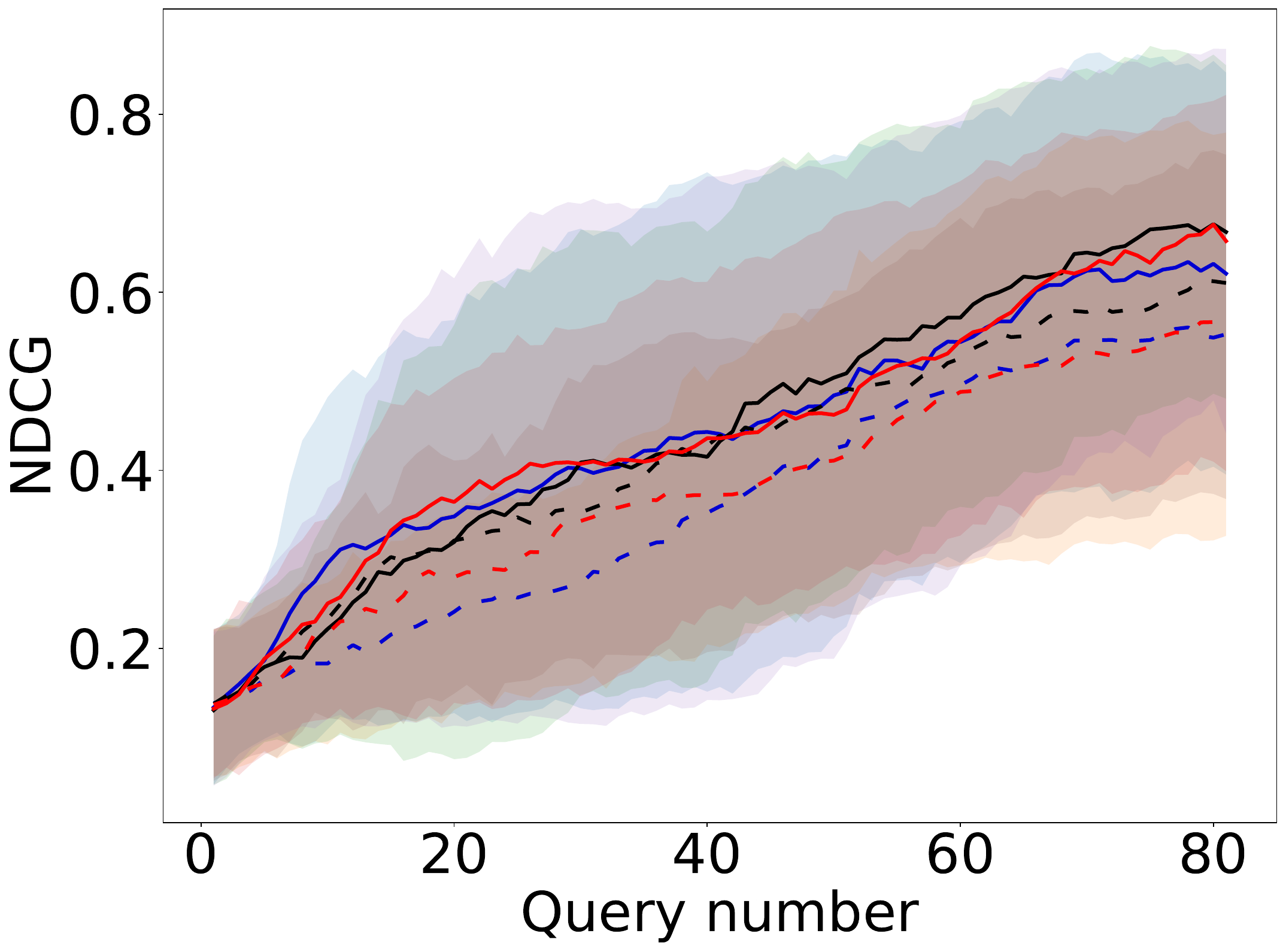}

    \caption{Batch vs.\ iterative posterior update.
    Left: RecSim NG. Right: Movielens}

    \label{fig:posterior_update}
\end{figure}

In Figure~\ref{fig:posterior_update}, we compare the batch and iterative posterior update methods (Section~\ref{subsec:param_posterior}.1). In both RecSim NG and MovieLens datasets, the iterative method outperforms batch across multiple optimization methods. Intuitively, while information from both the prior $P_U(u)$ and the query $q$ contribute to the posterior update, only the responses to $q$ provide new information. 
When the prior is far from the true user utility, especially at the start of training, the posterior update is dominated by information from $q$; thus, by updating the posterior iteratively, new information collected from each query response (in HMC generation of new posterior samples) compounds. As a result,
the iterative method generally has much better sample efficiency. Since user queries are expensive, we adopt the iterative method in all subsequent experiments.

\begin{figure}
    \centering
\includegraphics[width=0.5\textwidth]{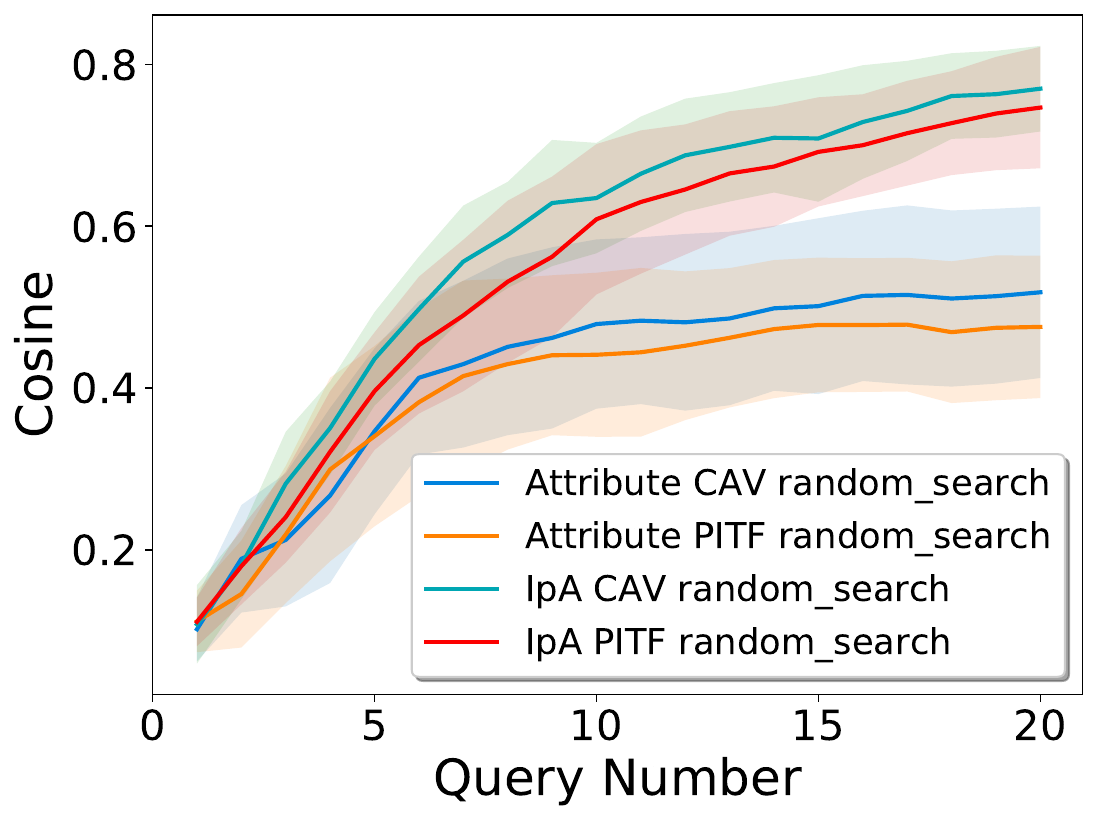}\,\includegraphics[width=0.5\textwidth]{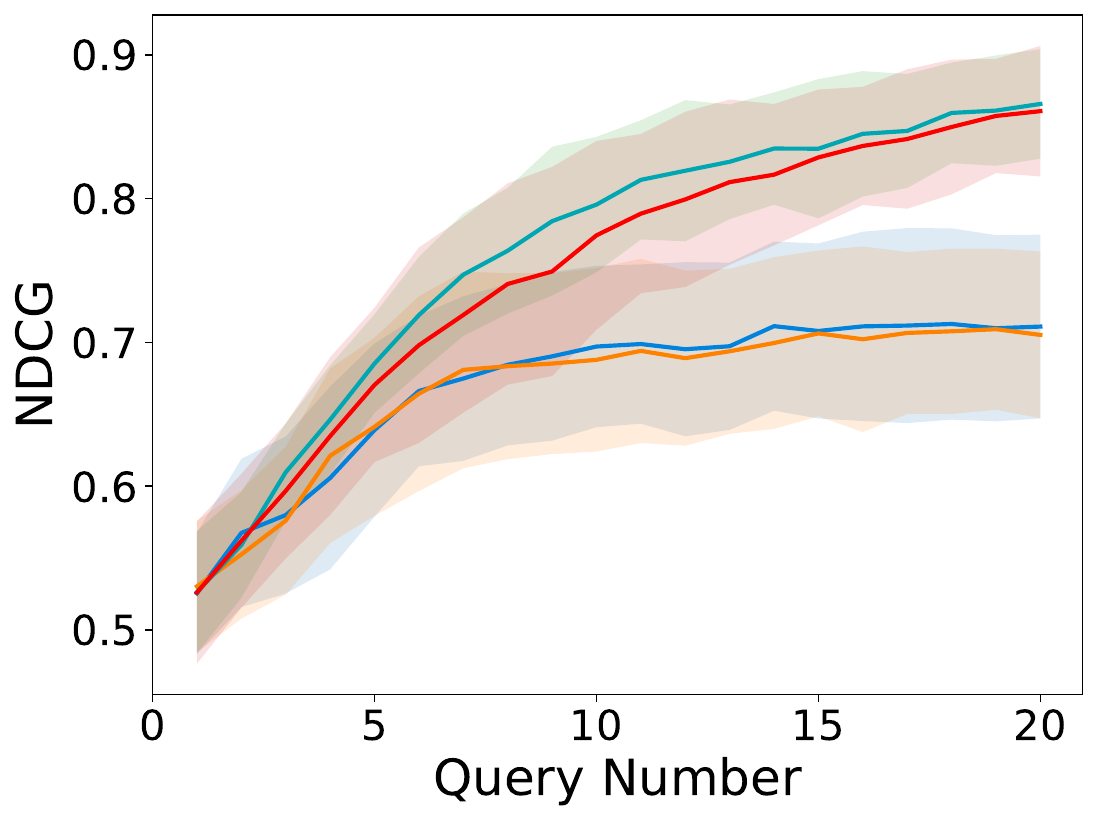}

\caption{CAV vs.\ PITF attribute semantics (RecSim NG).}

\label{fig:attr_binary_pitf}
\end{figure}

We next study the effect of using our CAV-based semantics on the performance of the PE algorithms. Our PE framework does not require using CAVs for soft attributes---it can work with any semantics discovery method that generates some form of $g$-score to quantify the degree to which an item $i$ satisfies a tag $g$. We compare \emph{PITF (pairwise interaction tensor factorization)}~\cite{rendle2010} as an alternative semantics. PITF is a tensor factorization method, originally developed for personalized tag prediction, which outputs a predicted tag $y_{u,i,g}$ for each user $u\in\calU$, item $i\in\calI$, and tag $g\in\calT$. To learn the PITF semantics, $\phi_g$, for a tag $g$, we fit a linear regressor to the dataset $\{(\phi_I(i), y_{u,i,g})\}_{u,i}$, where $u$ and $i$ are sampled from $\calU$ and $\calI$.\footnote{We use \texttt{https://github.com/yamaguchiyuto/pitf/} to train the PITF model.} In Figure~\ref{fig:attr_binary_pitf}, we compare the performance of PE algorithms that use the CAV and PITF representations with attribute and IpA queries, random search, and the EVOI AF. In all regimes, PE with CAVs performs better than with PITF, with a more significant advantage in cosine similarity (which reflects better ability to estimate the user's utility). This corroborates the main motivation for using CAVs, which is the ability to better represent semantic attributes that are especially predictive of users' preferences.

Finally, we consider the impact of CAV uncertainty on our PE algorithms. 
In contrast to the experiments above, we model the noise in CAV discovery by injecting various degrees of uncertainty into the CAV model $P_g(\phi_g | \bfD_g)$ for each tag $g$, where the more certain tags have lower (co-)variance in their CAV models. 
We assume $P_g$ is a multivariate normal distribution with mean $\mu_{g}$ and covariance matrix $\Sigma_{g}$. To explicitly model the fact that CAVs will various degrees of uncertainty, the mean of $P_g$ is the (deterministic) CAV learned via logistic regression, while the co-variance matrix is $\sigma_{g}^2 \calI$, where the standard deviation $\sigma_{g}$ is randomly chosen from a set of $|\mathcal T|$ values, ranging from $0.01$ to $1$ evenly spread on a $\log_{10}$ scale. For queries using attribute $g$, a ``true'' CAV vector $\phi_g$ is sampled from $P_g$ to power the user's responses (Eqs.~\ref{eq:BCAV_mean_slate}, \ref{eq:BCAV_mean_prob} and~\ref{eq:response_pref_CAV_query}).

\begin{figure}
    \centering
\includegraphics[width=0.5\textwidth]{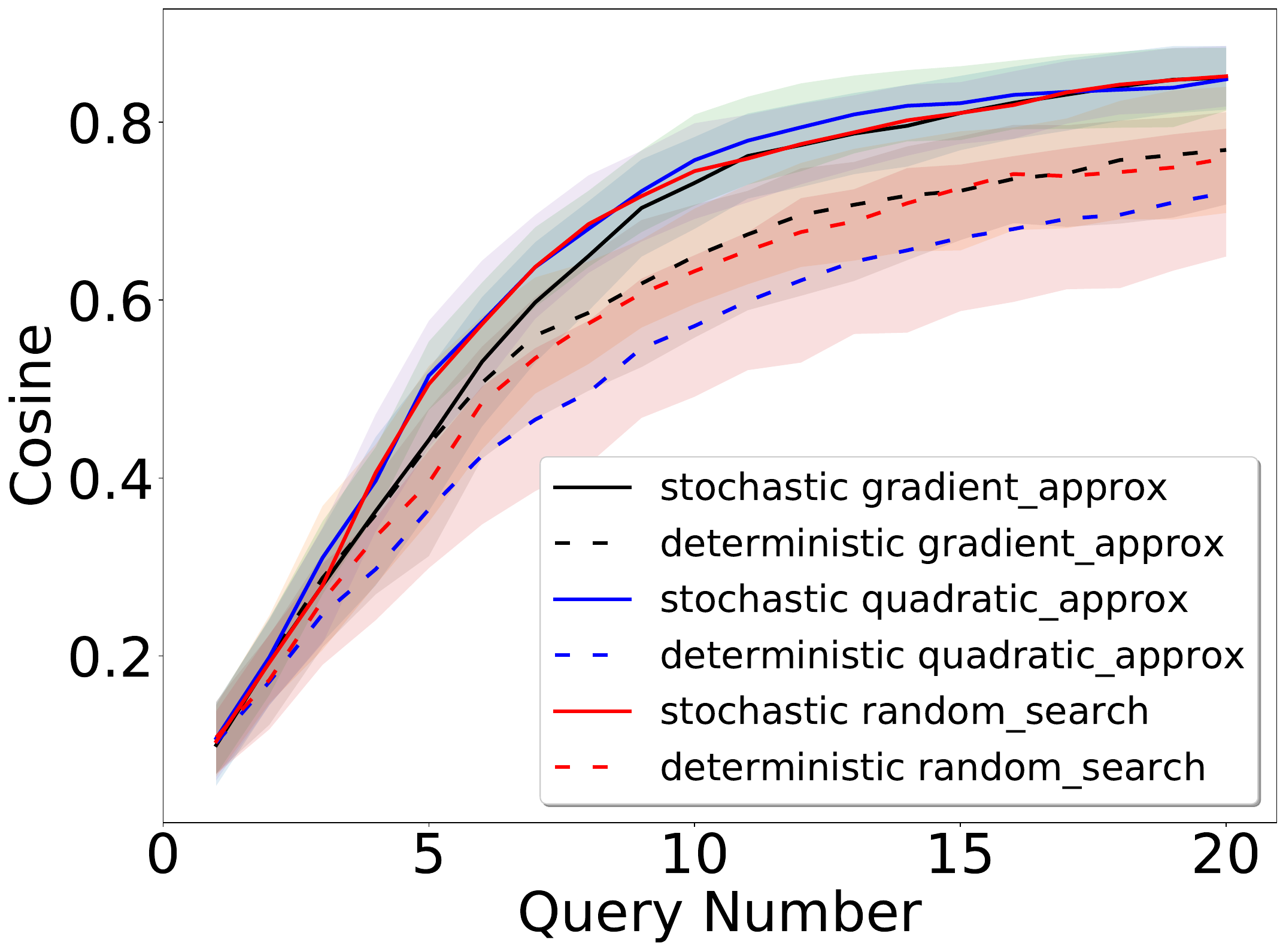}\,\includegraphics[width=0.5\textwidth]{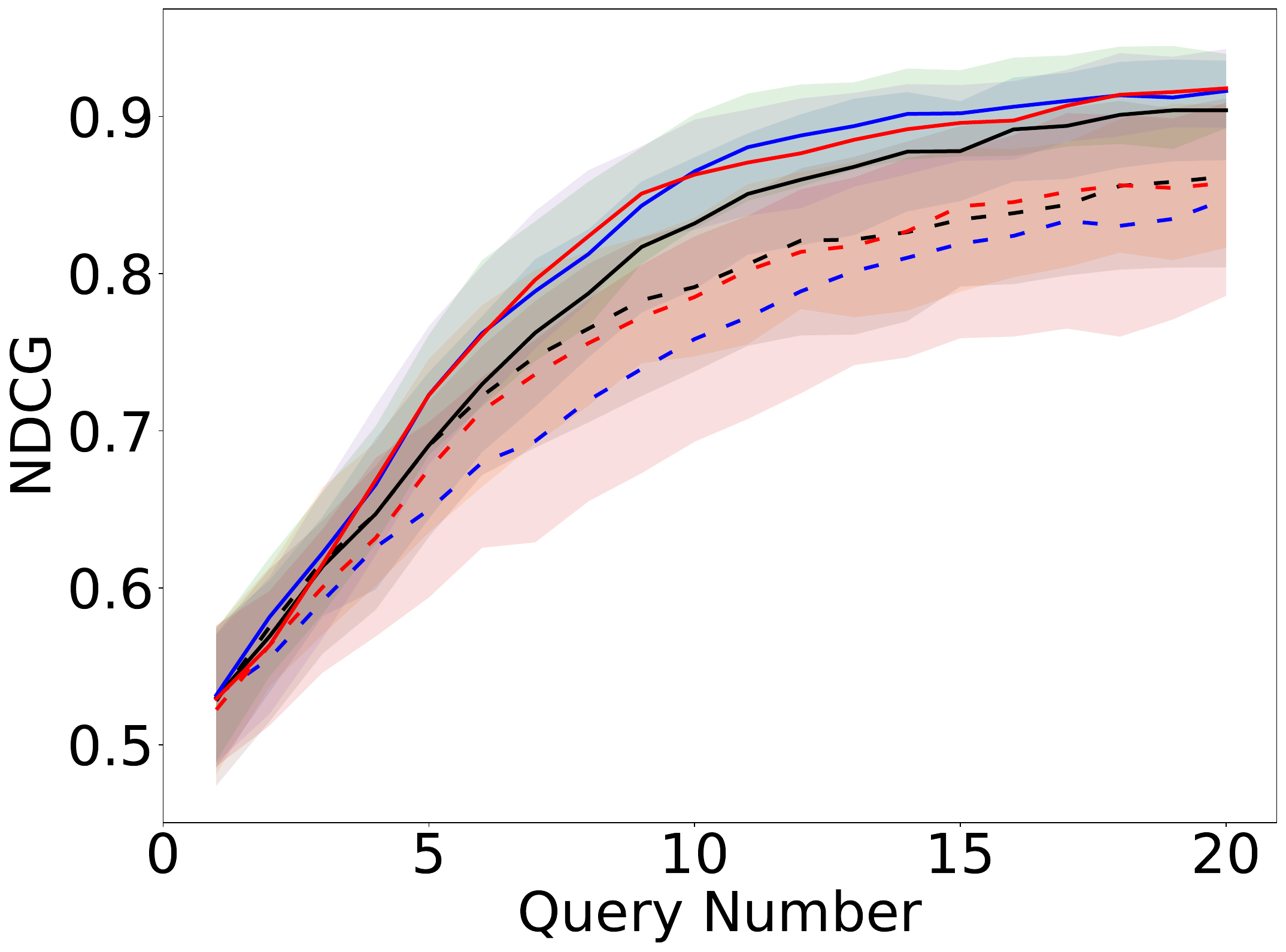}

\caption{PE with CAV uncertainty modeling (RecSim NG).}

\label{fig:synthetic_cav_uncertainty}
\end{figure}

We test whether modeling the CAV uncertainty in belief update and query optimization improves IG and RQ. Figure~\ref{fig:synthetic_cav_uncertainty} shows PE results of an experiment using IpA queries and three joint optimizers (random search, continuous relaxation with first- or second-order optimization). We see that modeling CAV uncertainty offers significant gain in IG and RQ, with a 10-15\% NDCG improvement over PE methods that update their beliefs by treating the mean CAV as ``certain''. Ignoring uncertainty (and CAV learning error) can generate over-confident error-prone belief updates.

\begin{figure}
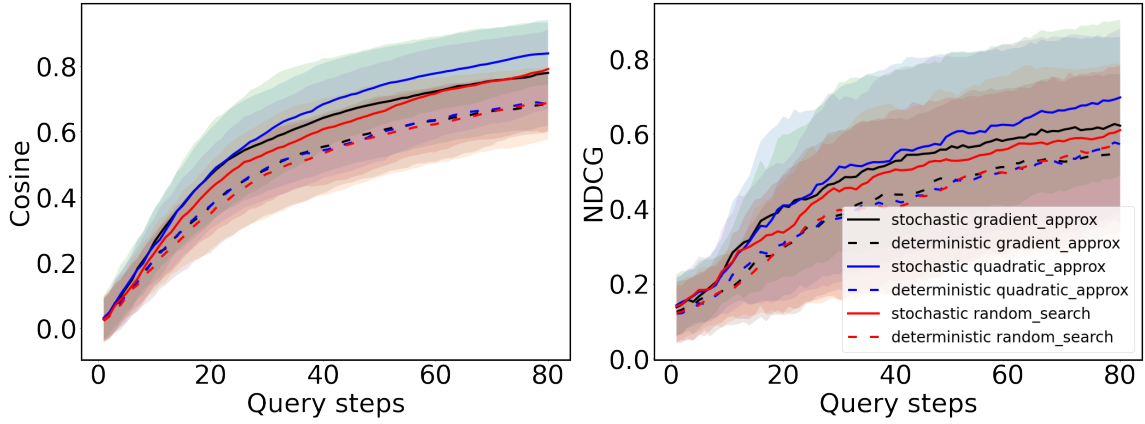

\centering
\includegraphics[width=0.5\textwidth]{figures/movielens_yinlam_cav_uncertainty_cosine_pdf.pdf}\,\includegraphics[width=0.5\textwidth]{figures/movielens_yinlam_cav_uncertainty_ndcg_pdf.pdf}

\caption{PE with CAV uncertainty modeling (MovieLens).}

\label{fig:movielens_cav_uncertainty}
\end{figure}

\begin{figure}
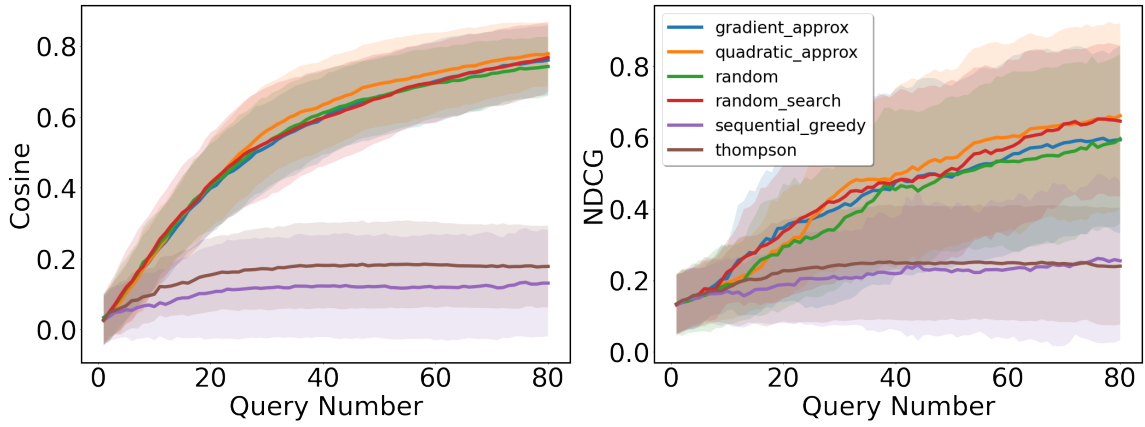

    \centering
\includegraphics[width=0.5\textwidth]{figures/movielens_yinlam_acquisition_method_cosine.pdf}\,\includegraphics[width=0.5\textwidth]{figures/movielens_yinlam_acquisition_method_ndcg.pdf}

    \caption{Comparing query optimizers (MovieLens).}

    \label{fig:movielens_acquisition_method}
\end{figure}

\begin{figure}
    \centering
\includegraphics[width=0.5\textwidth]{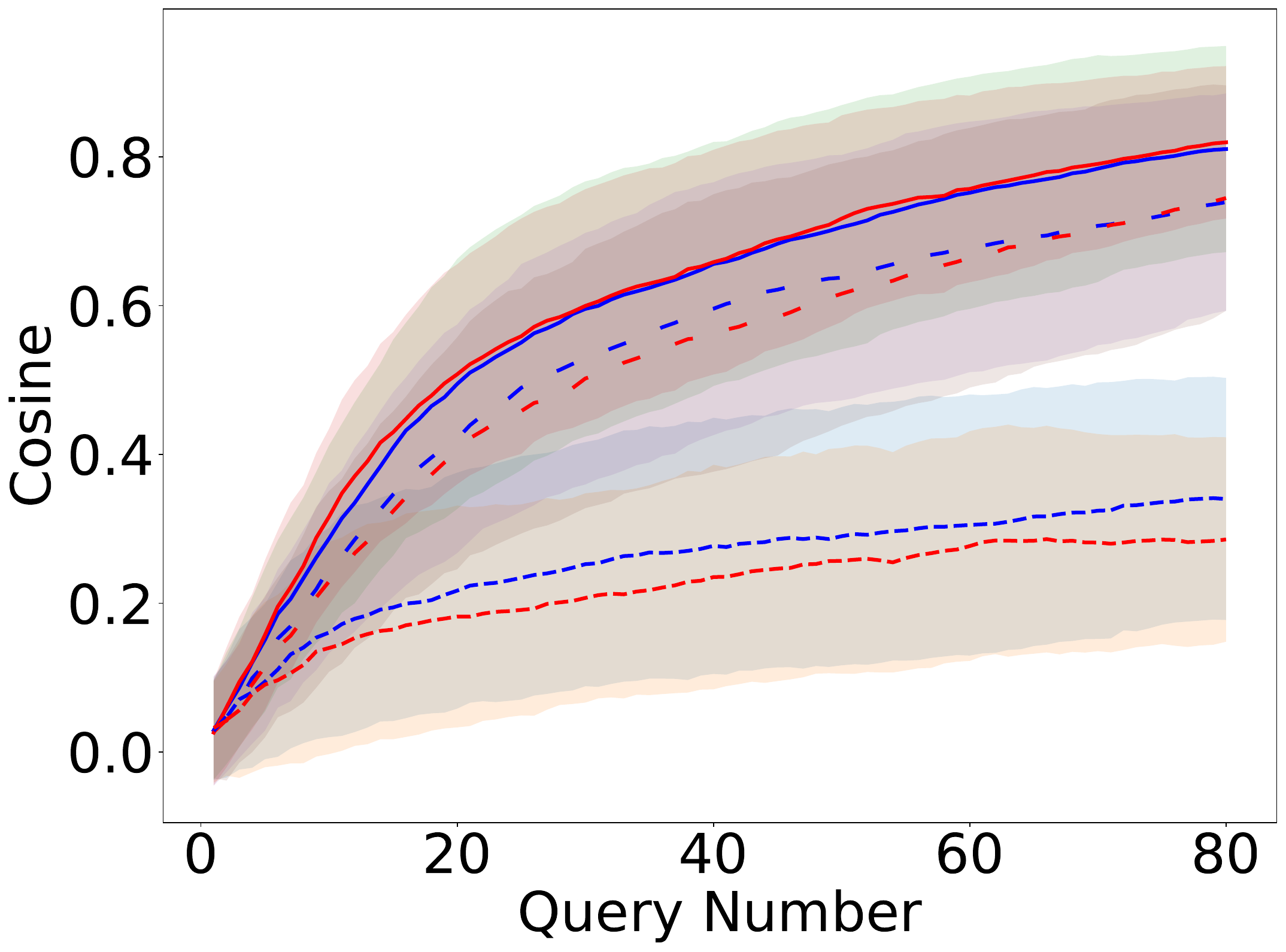}\,\includegraphics[width=0.5\textwidth]{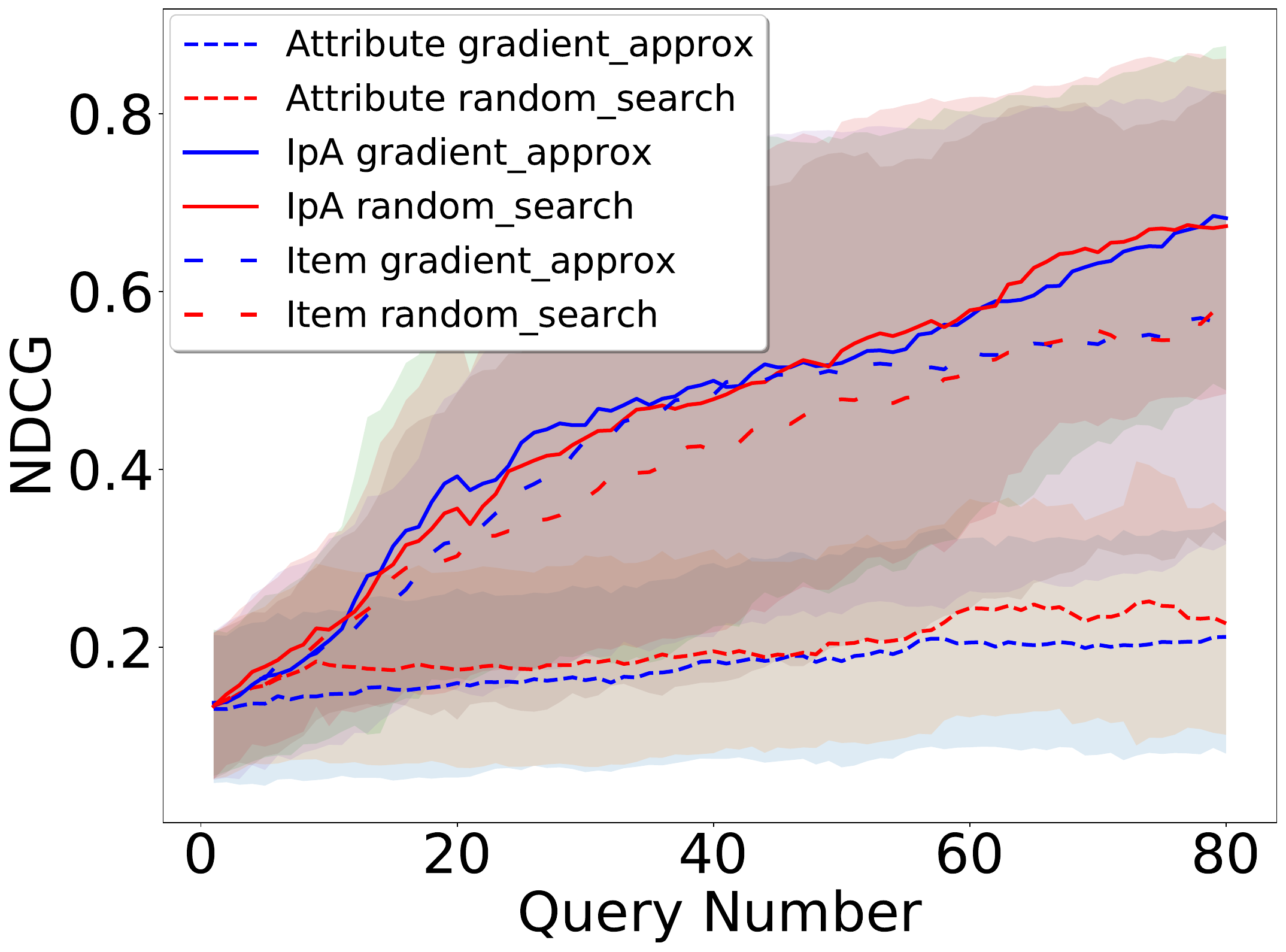}

    \caption{Comparing query types (MovieLens).}

    \label{fig:movielens_query_type}
\end{figure}



\noindent
\textbf{Experiment 4: MovieLens 20M.}
To evaluate PE with MovieLens 20M data, similar to the critiquing setup in \cite{gopfert2021discovering}, we create ``ground-truth'' users, each of whom has rated at least $50$ movies. Given the large number of ratings, their corresponding embeddings are relatively stable and accurate, so we treat them as the \emph{ground truth utility} for these users, which is used to generate query responses in our experiments. We sample $16$ such test users for PE. The RS belief state for each user is initialized with a \emph{cold-start prior} whose mean is the average embedding over all users. We use slates of $5$ movies, set 
$\gamma$ to $0.5$, and use EVOI AF and parameterized posteriors. 

We first evaluate the impact of CAV uncertainty.
Similar to the RecSim NG experiment above, for each attribute $g$ 
we assume the RS has a known multivariate-normal distribution $P_g$ with mean $\mu_{g}$ and covariance $\sigma_{g}^2 \calI$ that are generated as above to capture CAV uncertainty. 
We test how modeling the CAV uncertainty in belief update and query optimization impacts IG and RQ. Figure~\ref{fig:movielens_cav_uncertainty} shows PE results using IpA queries and three joint optimizers. Again, we see that modeling CAV uncertainty in PE improves both IG and RQ, with an up to 10--20\%  improvement in NDCG compared to PE using ``deterministic'' CAVs. This demonstrates the need to model noise in the estimation of the CAV semantics for soft attributes in PE, particularly in real-world domains as reflected in MovieLens. 

We next validate the performance of different query optimizers in the MovieLens setting. Figure~\ref{fig:movielens_acquisition_method} compares different optimizers w.r.t.\ IG and RQ. While the RS can learn about user preferences and improve its recommendations with most PE methods, the joint optimizers provide better recommendations (slates with greater NDCG across any number of queries) by extracting more useful information (greater cosine). As in RecSim NG, PE with random queries performs relatively well, likely due to two factors: (i) With an inaccurate prior model (e.g., during the early phases of PE), a random query of diverse items may effectively reduce belief-state entropy; (ii) With a combinatorial space of decision variables (slates), none of the optimization strategies converge to local optima (w.r.t.\ BPER) that are much better than the random query. This suggests that smartly switching between random and jointly optimized queries may  be ``practically optimal'' in terms of computational complexity and PE, but detailed studies are left for future work.    

Finally, we explore different query types with both random search and gradient approximation. Figure~\ref{fig:movielens_query_type} shows that PE with IpA queries is much more effective than PE with simpler queries, an observation consistent with that seen in the RecSim NG setting. With the additional information collected at each PE step, IpA queries offer a 10--20\% NDCG gain over item queries. PE with attribute queries performs worst w.r.t.\ both IG and RQ because, without allowing users to express nuanced preferences through item selection, it may fail to capture preferences regarding important but untaggable latent attributes.

\vspace{-0.1in}
\section{Conclusions and Future Work}
\label{sec:conclusion}

We have proposed a framework for preference elicitation in interactive recommenders that allows users to navigate item space or critique recommendations using soft attributes. Our techniques exploit concept-activation vectors (CAVs) to uncover the semantics of soft attributes w.r.t.\ the RS's item representation, and to support Bayesian updates of the RS's beliefs about a user's underlying preferences. 
We considered uncertainty in the semantics of soft attributes and develop scalable, continuous relaxation methods for query optimization.
We studied a variety of query types, both item and attribute-based, and response models, developed belief update and query optimization methods (i.e., acquisition functions) for each, and demonstrated the efficacy of these methods on both simulated and MovieLens 20M data. Our work has fundamental implications for designing the next-generation of recommenders (e.g., dialogue-based RSs), since our methods allows RSs to interact and learn about users more naturally with soft attributes. 


There are a number of avenues for future research. With insufficient tag data, CAV noise may render PE ineffective, requiring active learning methods for attribute semantics---integrating these into the PE process should prove valuable. \emph{Subjectivity}, where users may use terms in different ways \citep{gopfert2021discovering,radlinski_EtAl:aaai22}, means methods to elicit a user's ``personalized'' semantics for an attribute should further enhance PE. Multi-modal user belief model, reflecting more diverse preferences, are of interest. While our current PE methods optimize for myopic information gathering and recommendation, multi-step lookahead PE methods that use reinforcement learning remain an important challenge. Finally, experiments with real users are needed to further evaluate our methods with more diverse user preferences, attribute semantics, and responses.




\bibliographystyle{ACM-Reference-Format}
\bibliography{long,standard,tcav,refs}

\clearpage
\appendix


\section{Elicitation with CAV Uncertainty}
\label{app:CAVuncertainty}

\subsection{User Belief State Update and AFs}
To incorporate the CAV uncertainty, we consider a soft attribute query $\tilde{q}=(S, P_g)$, where RS has a CAV belief $P_g(\phi_g | \bfD_g)$ reflecting this uncertainty with training data $\bfD_g$. We assume the true CAV $\phi_g$ is sampled from $P_g$. In this work, we do not update the CAV belief with user responses to CAV semantic elicitation.
However, the RS has to compute the response probabilities for updating user belief state according to its CAV belief.

For a (soft) attribute query $\tilde{q}=(S, P_g)$, by referring to the user response model $P(\rho|q, \phi_u)$ we define 
\begin{align}\notag
    P(\rho \mid \tilde{q}, \phi_u)&= \EE_{\phi_g \sim P_g} P(\rho \mid (S, \phi_g), \phi_u) \\ 
    &=\int_{\phi_g } P_g(\phi_g | D_g) P(\rho \mid (S, \phi_g), \phi_u)d\phi_g
\end{align}
as the probability of observing response $\rho$ under a CAV belief $P_g$. We then update our belief about $u$ with responses:
$$P_U(u \mid \rho,\tilde{q}) := P(\phi_u \mid \rho,\tilde{q}) \propto P(\rho \mid \tilde{q},\phi_u) P_U(u).$$ 
Given any query-response pair $(\tilde{q},\rho)$ under the current belief state $P_U(u)$, the posterior belief is updated by Bayes rule:
\begin{equation}
    P(\rho \mid \tilde{q}, \phi_u) = \frac{P(\rho \mid \tilde{q},\phi_u) P(\phi_u)}{\int_{\phi_u} P(\rho \mid \tilde{q},\phi_u) P(\phi_u) d\phi_u}.
\end{equation}

We also need to consider the CAV uncertainty in computing acquisition functions. For example, the EVOI of $\tilde{q}$ (given $\calH$) is:
\begin{equation}
\EVOI(\tilde{q} \mid \calH) = \PEU(\tilde{q} \mid \calH) - \EU^*\big(P_U(u \mid \calH)\big).
\end{equation}
where $EU^*(P_U(u))$ is still the same as Eq (\ref{eq:eu_definition})
but $PEU(\tilde{q} \mid \calH)$ is the \emph{posterior expected utility} of w.r.t.\ $\tilde{q}$,
\begin{equation}
PEU(\tilde{q} \mid \calH) = \sum_{\rho} P(\rho \mid \tilde{q},\calH) \cdot EU^*\big(P_U(u \mid \calH\cup\{(\rho,\tilde{q})\})\big).
\end{equation}
Other acquisition functions can also be extended analogously but their details will be omitted for the sake of brevity. 

\subsection{Continuous Relaxation for AFs}\label{app:relaxation}

Recall that the acquisition function can be expressed as

\begin{equation}
    F(q)=\gamma \IG(q\mid\calH) + (1 - \gamma)\RQ(q\mid\calH),
\end{equation}
where
\begin{equation}
    \RQ(q \mid \calH):=\sum_{i\in S}\mathbb{E}_{\phi_u\sim P_U(u\mid\calH)}[\phi_u^\top\phi_I(i)]
\end{equation}
measures the recommendation quality and $\IG$ is the information gathering metric (i.e., Entropy, MI or EVOI). In the following we take EVOI as an example and show how we obtain an differentiable objective function. The same derivation applies to Entropy and MI. For IG=EVOI, computing $q^*=\arg\max_{q} F(q)$ is equivalent to finding

\begin{equation}
    q^*=\arg\max_{q}\big\{ \gamma PEU(q|\calH) +(1-\gamma)RQ(q|\calH)\big\}.
\end{equation}

PEU
requires computing the expected belief w.r.t.\ possible responses $\rho$, $P_U(u\mid\calH\cup\{(\rho,q)\})$. This computation can be expensive, so we approximate PEU by sampling from the prior $P_U(u\mid\calH)$:
\begin{equation}
\label{eq:approx_peu_compute}
\!\!\!\!PEU'(q \mid \calH) \!:=\! \sum_{\rho} \max_{i\in \calI} \mathbb{E}_{\phi_u \sim P_U(u \mid \calH)}\big[\phi_u^\top\phi_I(i)\cdot P(\rho \mid q,\phi_u)\big].
\end{equation}
With CAV uncertainty, the response probabilities used in PEU and other AFs are computed using expectation over CAV. The PEU function in Eq \eqref{eq:approx_peu_compute} can be further approximated by drawing $m$ samples $\{\phi_{u,j}\}_{j=1}^m$  from the posterior $P_U(u|\calH)$:
\begin{equation}\label{eq:approx_peu_2}
\!\!\!\!PEU(q \mid \calH) \approx \frac{1}{m} \sum_{\rho} \max_{i\in \calI} \big[\sum_{j=1}^m\phi_{u,j}^\top\phi_I(i) P(\rho \mid q,\phi_{u,j})\big].
\end{equation}
And similarly, the $RQ$ function can be approximated by
\begin{equation}
\!\!\!\!RQ(q \mid \calH) \approx \frac{1}{m} \sum_{i\in S}\big[\sum_{j=1}^m\phi_{u,j}^\top\phi_I(i) \big].
\end{equation}

In the following we use the concatenation of attribute vector $g$ and item embeddings in $S$ to represent $q$, immediately $RQ(q \mid \calH)$ is differentiable w.r.t. $q$. However, computing $PEU(q \mid \calH)$ requires enumerating over $\calI$ and thus is not differentiable. To derive a differentiable objective function $F(q)$, we assume $\calI$ is rich enough such that for each query-response pair $(q, \rho)$ we can find $i^*_{(q, \rho)} = \frac{\sum_{j=1}^m \phi_{u,j} P(\rho \mid q,\phi_{u,j})}{\|\sum_{j=1}^m \phi_{u,j} P(\rho \mid q,\phi_{u,j})\|} \max_{i\in\calI} \|\phi_I(i)\|$ to maximize the RHS of Eq \eqref{eq:approx_peu_2}. Then Eq \eqref{eq:approx_peu_2} can be further approximated by 
    \begin{equation}\label{eq:approx_peu_max}
            PEU(q \mid \calH)\! \approx \!\frac{ \max_{i\in\calI} \|\phi_I(i)\|}{m}\sum_{\rho} \Big\|\sum_{j=1}^m \phi_{u,j} P(\rho \mid q,\phi_{u,j})\Big\|_2,
    \end{equation}
which is differentiable w.r.t. $q$ because the response model $P(\cdot|q,\cdot)$ is differentiable w.r.t. $q$. 

Now we extend the continuous relaxation framework to account for queries with an uncertain CAV vector as its attribute. We may apply the re-parameterization trick to derive a continuous acquisition function. Suppose the CAV vector $\phi_g$ in the attribute query $\tilde{q}=(S, P_g)$ follows a Multivariate normal $\N(\mu_g, \Sigma_g)$. Then we can sample $n$ queries $\{q_i|q_i=(\{\phi_I(i)\}_{i\in S}, \phi_{g,i})\}_{i=1}^n$ by calculating $\phi_{g,i} = \mu_g + L_g \epsilon_i$, where $\Sigma_g = L_g^\top L_g$ is the Cholesky decomposition of the covariance matrix, and $\epsilon_i \sim \N(0, \I_d)$ are i.i.d. standard normal random vectors. Based on Eq \eqref{eq:approx_peu_2}, we can now introduce the uncertainty of CAVs in the following approximation of PEU

\begin{equation}\notag
        PEU(\tilde{q} \mid \calH) \approx \frac{ \max_{i\in\calI} \|\phi_I(i)\|}{mn}\sum_{i=1}^n\sum_{\rho} \Big\|\sum_{j=1}^m \phi_{u,j} P(\rho \mid q_i,\phi_{u,j})\Big\|_2,
\end{equation}
where 
\begin{align}\notag
    P(\rho=+1 \mid q_i,\phi_u) &= P(\rho=+1 \mid S,\mu_g,\Sigma_g,\epsilon_i,\phi_u) \\ \label{eq:response_P_continuous} & = \Phi\Big(\frac{(\mu_g + L_g \epsilon_i)^{\top}(\phi^*_{I,u}-\phi_{I,\bar{S}})}{\sigma_g}\Big),
\end{align}
Eq.~\eqref{eq:response_P_continuous} is differentiable w.r.t. $q=(\{\phi_I(i)\}_{i\in S},\mu_g, L_g)$ and we can thus obtain its gradient.
Suppose $(\hat{\phi_i}_{i\in S}, \hat{\mu}_g, \hat{L}_g)$ is a solution of the corresponding continuous optimization problem. The projection is done by picking the attribute with the smallest KL-divergence to $\N(\hat{\mu}_g, \hat{L}_g^{\top}\hat{L}_g)$ and the slate of items with the minimum Euclidean distance to $\hat{\phi_i}_{i\in S}$. 

\subsection{Gradient-based Query Optimization}\label{app:gradient_of_acquisition_wrt_slate}
Once we approximate the acquisition function in a continuous form $F(q)$, standard gradient-based method can be applied to find the optimal query $q*=\arg\max_q F(q)$. In our empirical study we use both first and second-order optimization. First we generate an initial query $q_0$ from random search with 20 trials and then apply either the following updates for a fixed number of steps:
\begin{enumerate}
    \item First-order optimization: $$q_t = q_{t-1} + \eta \nabla F(q_{t-1}), 1\leq t\leq T,$$
    \item Second-order optimization: $$q_t = q_{t-1} + [\lambda I +\nabla^2 F(q_{t-1})]^{-1}\nabla F(q_{t-1}), 1\leq t\leq T,$$
\end{enumerate}
where $\eta$ is the learning rate, $\lambda$ is a regularization constant used to guarantee the non-singularity of the Hessian matrix $\nabla^2 F$. In our experiments we take $T=2, \eta=1e-3, \lambda=1e-4$. 

\commentout{
\section{Gradient of Acquisition Functions with respect to Query Slate}
\subsection{Mutual Information w.r.t. Attribute Queries with Mean Slate Model}
Now, we want to compute the following expression:
\begin{align}
    \frac{\partial MI(\phi_{I,u}^* ; \rho^{(k)} \mid q^{(k)},\mathcal H^{(k-1)})}{\partial \phi_{I,\bar{S}^{(k)}}}
\end{align}
where $q^{(k)}$ is a binary CAV query with the mean slate model. To compute this gradient, let's first expand the mutual information as:
\begin{align}
    H(\rho^{(k)} \!\mid\! q^{(k)}, \mathcal H^{(k-1)}) \!-\! \mathbb{E}_{\phi_u \sim P_U \mid \mathcal H^{(k-1)}}\left[H(\rho^{(k)} \!\mid\! q^{(k)}, \phi_u)\right]
\end{align}
We can write the first entropy term as:
\begin{align}
    &-\sum_{\rho^{(k)}}P(\rho^{(k)} \mid q^{(k)}, \mathcal H^{(k-1)})\log P(\rho^{(k)} \mid q^{(k)}, \mathcal H^{(k-1)})\nonumber\\
    &= -\sum_{\rho^{(k)}}\!\bigg[\mathbb{E}_{\phi_u\sim P_U \mid \mathcal H^{(k-1)}}\left[P(\rho^{(k)} \mid q^{(k)}, \phi_u)\right] \times \nonumber\\
    &\qquad\qquad \log \mathbb{E}_{\phi_u \sim P_U \mid \mathcal H^{(k-1)}}\left[P(\rho^{(k)} \mid q^{(k)}, \phi_u)\right]\bigg]\nonumber\\
    &= -\sum_{\rho^{(k)}}\Bigg[\mathbb{E}_{\phi_u\sim P_U \mid \mathcal H^{(k-1)}}\left[\Phi\left(\rho^{(k)}\frac{\phi_{g^{(k)}}^\top (\phi_{I,u}^* \!-\! \phi_{I,\bar{S}^{(k)}})}{\sigma_{g^{(k)}}}\right)\right]\times \nonumber\\
    &\qquad \log \mathbb{E}_{\phi_u\sim P_U \mid \mathcal H^{(k-1)}}\!\left[\Phi\left(\rho^{(k)}\frac{\phi_{g^{(k)}} \cdot (\phi_{I,u}^* \!-\! \phi_{I,\bar{S}^{(k)}})}{\sigma_{g^{(k)}}}\right)\right]\Bigg]\:.
\end{align}
Similarly, we can write the second entropy term as:
\begin{align}
    &-\mathbb{E}_{\phi_u\sim P_U \mid \mathcal H^{(k-1)}}\left[\sum_{\rho^{(k)}}P(\rho^{(k)} \mid q^{(k)}, \phi_u)\log P(\rho^{(k)} \mid q^{(k)}, \phi_u)\right]\nonumber\\
    &= -\mathbb{E}_{\phi_u\sim P_U \mid \mathcal H^{(k-1)}}\Bigg[\sum_{\rho^{(k)}}\Phi\bigg(\rho^{(k)}\frac{\phi_{g^{(k)}}^\top (\phi_{I,u}^* - \phi_{I,\bar{S}^{(k)}})}{\sigma_{g^{(k)}}}\bigg) \nonumber\\
    &\qquad\log \Phi\bigg(\rho^{(k)}\frac{\phi_{g^{(k)}}^\top (\phi_{I,u}^* - \phi_{I,\bar{S}^{(k)}})}{\sigma_{g^{(k)}}}\bigg)\Bigg]\:.
\end{align}

Next, we calculate the gradients of these terms with respect to $\phi_{I,\bar{S}^{(k)}}$. The gradient of the first entropy term is:
\begin{align}
    &\sum_{\rho^{(k)}}\mathbb{E}_{\phi_u\sim P_U \mid \mathcal H^{(k-1)}}\left[\varphi\left(\rho^{(k)}\frac{\phi_{g^{(k)}}^\top (\phi_{I,u}^* - \phi_{I,\bar{S}^{(k)}})}{\sigma_{g^{(k)}}}\right)\frac{\rho^{(k)}\phi_{g^{(k)}}}{\sigma_{g^{(k)}}}\right] \nonumber\\
    &\qquad\left(1\!+\!\log \mathbb{E}_{\phi_u\sim P_U \mid \mathcal H^{(k-1)}}\!\left[\Phi\left(\rho^{(k)}\frac{\phi_{g^{(k)}}^\top \!(\phi_{I,u}^* \!-\! \phi_{I,\bar{S}^{(k)}})}{\sigma_{g^{(k)}}}\!\right)\right]\right)\,.
\end{align}
Similarly, the gradient of the second entropy term is:
\begin{align}
    &\mathbb{E}_{\phi_u \sim P_U \mid \mathcal H^{(k-1)}}\Bigg[\sum_{\rho^{(k)}}\varphi\left(\rho^{(k)}\frac{\phi_{g^{(k)}}^\top (\phi_{I,u}^* - \phi_{I,\bar{S}^{(k)}})}{\sigma_{g^{(k)}}}\right)\frac{q^{(k)}c_{g^{(k)}}}{\sigma_{g^{(k)}}}\nonumber\\
    &\qquad\left(1+\log \Phi\left(\rho^{(k)}\frac{\phi_{g^{(k)}}^\top (\phi_{I,u}^* - \phi_{I,\bar{S}^{(k)}})}{\sigma_{g^{(k)}}}\right)\right)\Bigg]\:.
\end{align}
Together, these two expression are sufficient to compute the overall gradient.

\subsection{EVOI w.r.t. Attribute Queries with Mean Slate Model}
Finally, we want to compute the gradient of EVOI with respect to an attribute query with the mean slate model. Mathematically, we want to compute:
\begin{align}
    \frac{\partial EVOI(q^{(k)} \mid \mathcal H^{(k-1)})}{\partial \phi_{I,\bar{S}^{(k)}}}
\end{align}
Since the second term in EVOI (see Eq.~\eqref{eq:evoi_definition}) does not depend on the query, we can expand this expression as:
\begin{align}
    \frac{\partial \sum_{\rho^{(k)}} \max_{i\in \calI} \mathbb{E}_{\phi_u\sim P_U \mid \mathcal H^{(k-1)}}\left[\phi_u^\top\phi_I(i)P(\rho^{(k)}\mid q^{(k)},\phi_u)\right]}{\partial \phi_{I,\bar{S}^{(k)}}}
\end{align}
To further simplify, we first define:
\begin{align}
    i_{\rho^{(k)}} := \argmax_{i\in\calI} \mathbb{E}_{\phi_u\sim P_U \mid \mathcal H^{(k-1)}}\left[\phi_u^\top\phi_I(i)P(\rho^{(k)}\mid q^{(k)},\phi_u)\right]
\end{align}
Then, the gradient can be written as follows:
\begin{align}
    &-\sum_{\rho^{(k)}} \mathbb{E}_{\phi_u\sim P_U \mid \mathcal H^{(k-1)}}\Bigg[\phi_u^\top\phi_I(i_{\rho^{(k)}}) \nonumber\\
    &\qquad \varphi\left(\rho^{(k)}\frac{\phi_{g^{(k)}}^\top (\phi_{I,u}^* - \phi_{I,\bar{S}^{(k)}})}{\sigma_{g^{(k)}}}\right)\frac{\rho^{(k)}\phi_{g^{(k)}}}{\sigma_{g^{(k)}}}\Bigg]\:.
\end{align}
}

\section{Gradient of Log-Posterior}
\label{app:gradient_of_logposterior}

We are interested in the gradient of the following expression with respect to $\phi_u$ (the embedding sample for the user):
\begin{align}
    -\frac12 & (\phi_u - \phi_{\mu,U}(u))^\top {\left(\phi_{\sigma,U}(u)^\top\phi_{\sigma,U}(u)\right)}^{-1}(\phi_u - \phi_{\mu,U}(u)) + \nonumber\\
    &\sum_{k=1}^{K} \log P(\rho^{(k)} \mid q^{(k)}, \phi_u)
\end{align}
For the first term that comes from the prior, the gradient is equal to $-\left(\phi_{\sigma,U}(u)^\top\phi_{\sigma,U}(u)\right)(\phi_u - \phi_{\mu,U}(u))$. The components of the second term (each log likelihood) depends on the type of query. 

\subsection{Attribute Queries}
\subsubsection{Mean Slate Model}
Here, we have
\begin{align}
    \log P(\rho^{(k)} \mid q^{(k)},  \phi_u) = \log \Phi\left(\rho^{(k)}\frac{\phi_{g^{(k)}}^\top (\phi_{I,u}^* - \phi_{I,\bar{S}^{(k)}})}{\sigma_{g^{(k)}}}\right)\:,
\end{align}
where $\rho^{(k)}$ is either $-1$ or $1$. We can write the gradient of this expression with respect to $\phi_u$ as follows by using the relation between $\phi_u$ and $\phi_{I,u}^*$:
\begin{align}
    \frac{\varphi\left(\rho^{(k)}\frac{\phi_{g^{(k)}}^\top (\phi_{I,u}^* - \phi_{I,\bar{S}^{(k)}})}{\sigma_{g^{(k)}}}\right)}{\Phi\left(\rho^{(k)}\frac{\phi_{g^{(k)}}^\top (\phi_{I,u}^* - \phi_{I,\bar{S}^{(k)}})}{\sigma_{g^{(k)}}}\right)} \! \frac{\rho^{(k)}z}{\sigma_{g^{(k)}}} \! \left(\frac{\phi_{g^{(k)}}}{\|\phi_u\|_2} \!-\! \frac{\phi_{g^{(k)}}^\top (\phi_{I,u}^* \!-\! \phi_{I,\bar{S}^{(k)}})\phi_u}{\|\phi_u\|_2^3}\right)\,,  
\end{align}
where $\varphi$ is the standard normal pdf, and $z=\max_{i\in\mathcal I} \|\phi_I(i)\|_2$.

\subsubsection{Mean Probability Model}
We start with
\begin{align}
    \log P(\rho^{(k)} \!\mid\! q^{(k)},  \phi_u) \!=\! \log \frac{1}{\lvert S^{(k)}\rvert}\sum_{i\in S^{(k)}}\!\Phi\left(\rho^{(k)}\frac{\phi_{g^{(k)}}^\top (\phi_{I,u}^* \!-\! \phi_I(i))}{\sigma_{g^{(k)}}}\right)\,,
\end{align}
where $\rho^{(k)}$ is either $-1$ or $1$. Again by using the relation between $\phi_u$ and $\phi_{I,u}^*$, we can write the gradient of this expression with respect to $\phi_u$ as follows:
\begin{align}
    \frac{\sum\limits_{i\in S^{(k)}}\!\varphi\left(\rho^{(k)}\frac{\phi_{g^{(k)}}^\top (\phi_{I,u}^* - \phi_I(i))}{\sigma_{g^{(k)}}}\right)\!\frac{\rho^{(k)}z}{\sigma_{g^{(k)}}}\!\left(\frac{\phi_{g^{(k)}}}{\|\phi_u\|_2}\!-\!\frac{\phi_{g^{(k)}}^\top (\phi_{I,u}^* - \phi_I(i))\phi_u}{\|\phi_u\|_2^3}\right)}{\sum_{i\in S^{(k)}}\Phi\left(\rho^{(k)}\frac{\phi_{g^{(k)}}^\top (\phi_{I,u}^* - \phi_I(i))}{\sigma_{g^{(k)}}}\right)}\:.
\end{align}

\subsection{Item Queries}
We ignore the temperature $T$ for simplicity. Here, we have
\begin{align}
   & \log P(\rho^{(k)} \mid q^{(k)},  \phi_u) = \log \frac{\exp(\phi_u^\top\phi_I(\rho^{(k)}))}{\sum_{i\in S^{(k)}}\exp(\phi_u^\top\phi_I(i))} \nonumber\\
    &= \phi_u^\top\phi_I(\rho^{(k)}) - \log\sum_{i\in S^{(k)}}\exp(\phi_u^\top\phi_I(i))\:,
\end{align}
where $\rho^{(k)}$ is an item from $S^{(k)}$. We can write the gradient of this expression with respect to $\phi_u$ as follows:
\begin{align}
    \phi_I(\rho^{(k)})-\frac{\sum_{i\in S^{(k)}}\exp(\phi_u^\top\phi_I(i))\phi_I(i)}{\sum_{i\in S^{(k)}}\exp(\phi_u^\top\phi_I(i))}\:.
\end{align}

\subsection{Item-Plus-Attribute Queries}
We again start with the response model:
\begin{align}
    &\log P(\rho^{(k)} \mid q^{(k)}, \phi_u) \nonumber\\
    &= \log \frac{\exp(\phi_u^\top\phi_I(\rho_1^{(k)}))}{\sum\limits_{i\in S^{(k)}}\exp(\phi_u^\top\phi_I(i))} \!+\! \log\Phi\left(\rho_2^{(k)}\frac{\phi_{g^{(k)}}^\top (\phi_{I,u}^* \!-\! \phi_I(\rho_1^{(k)}))}{\sigma_{g^{(k)}}}\right)\:,
\end{align}
where $\rho_1^{(k)}\in S^{(k)}$ and $\rho_2^{(k)}\in\{-1,+1\}$. We can write the gradient of this expression with respect to $\phi_u$ simply as a sum of the two gradients we derived before:
\begin{align}
    \phi_I&(\rho_1^{(k)}) \!-\! \frac{\sum\limits_{i\in S^{(k)}}\!\exp(\phi_u^\top\phi_I(i))\phi_I(i)}{\sum\limits_{i\in S^{(k)}}\!\exp(\phi_u^\top\phi_I(i))} \!+\! \frac{\varphi\!\left(\rho_2^{(k)}\frac{\phi_{g^{(k)}}^\top (\phi_{I,u}^* - \phi_I(\rho_1^{(k)}))}{\sigma_{g^{(k)}}}\right)}{\Phi\!\left(\rho_2^{(k)}\frac{\phi_{g^{(k)}}^\top (\phi_{I,u}^* - \phi_I(\rho_1^{(k)}))}{\sigma_{g^{(k)}}}\right)} \!\times \nonumber\\
    & \frac{\rho_2^{(k)}z}{\sigma_{g^{(k)}}} \! \left(\frac{\phi_{g^{(k)}}}{\|\phi_u\|_2} \!-\! \frac{\phi_{g^{(k)}}^\top (\phi_{I,u}^* \!-\! \phi_I(\rho_2^{(k)}))\phi_u}{\|\phi_u\|_2^3}\right)\,.
\end{align}

\commentout{
\section{Derivation of Laplacian Posterior Covariance}\label{app:H_derivation}
Similar to Appendix~\ref{app:gradient_of_logposterior}, the Hessian depends on the type of query $q^{(k)}$. Hence, we separately derive the two query types we implemented: attribute queries with the mean slate model, and item queries.

\subsection{Attribute Queries}
\subsubsection{Mean Slate Model}
We are interested in computing
\begin{align*}
	\mathbf{H} = -\left[\frac{\partial^2 \sum_{k=1}^{K} \log \Phi\left(\rho^{(k)}\frac{\phi_{g^{(k)}}^\top (\phi_{I,u}^* - \phi_{I,\bar{S}^{(k)}})}{\sigma_{g^{(k)}}}\right) }{{\partial \phi_u}^2}\right]_{\phi_u=\hat{\phi}_{\mu,U}(u)}\:.
\end{align*}
First, let
\begin{align}
    w_{k} \!=\! \rho^{(k)}\frac{\phi_{g^{(k)}}^\top (\phi_{I,u}^* \!-\! \phi_{I,\bar{S}^{(k)}})}{\sigma_{g^{(k)}}} \!=\! \rho^{(k)}\frac{\phi_{g^{(k)}}^\top \!\left(\frac{z\phi_u}{\left({\phi_u}^\top{\phi_u}\right)^{1/2}} \!-\! \phi_{I,\bar{S}^{(k)}}\right)}{\sigma_{g^{(k)}}}\:.
\end{align}
Also let $w'_{k}$ and $w''_{k}$ denote the derivative and the Hessian of $w_{k}$ with respect to $\phi_u$:
\begin{align}
    w'_{k} &= \rho^{(k)}\frac{z \phi_{g^{(k)}}}{\sigma_{g^{(k)}}({\phi_u}^\top {\phi_u})^{1/2}} - \rho^{(k)}\frac{z \phi_{g^{(k)}}^\top {\phi_u} {\phi_u}}{\sigma_{g^{(k')}} ({\phi_u}^\top {\phi_u})^{3/2}}\:,\\
    w''_{k} &= -\rho^{(k)}\frac{z \phi_{g^{(k)}} {\phi_u}^\top}{\sigma_{g^{(k)}} ({\phi_u}^\top {\phi_u})^{3/2}} - \rho^{(k)}\frac{z {\psi} \phi_{g^{(k)}}^\top + z \phi_{g^{(k)}}^\top {\phi_u} I_d}{\sigma_{g^{(k)}}({\phi_u}^\top {\phi_u})^{3/2}} + \nonumber\\
    &\qquad \rho^{(k)}\frac{3 z \phi_{g^{(k)}}^\top {\phi_u} {\phi_u} {\phi_u}^\top}{\sigma_{g^{(k)}}({\phi_u}^\top {\phi_u})^{5/2}}\:.
\end{align}
Then, we can write the expression for $\mathbf{H}$ as:
\begin{align}
    -\!\left[\sum_{k=1}^{K} \!\left( \frac{\varphi'(w_{k}) w'_{k} \Phi(w_{k}) \!-\! \varphi(w_{k})^2 w'_{k}}{\Phi(w_{k})^2}{w'_{k}}^\top \!+\! \frac{\varphi(w_{k})}{\Phi(w_{k})}w''_{k'} \right)\right]_{\phi_u=\hat{\phi}_{\mu,U}(u)}
\end{align}
where $\varphi'$ is the derivative of the standard normal's pdf, i.e., $\varphi'(x) = -x\varphi(x)$.

\subsection{Item Queries}
We want to compute
\begin{align*}
	\mathbf{H} = -\left[\frac{\partial^2 \sum_{k=1}^{K} \log \frac{\exp(\phi_u^\top\phi_I(\rho^{(k)}))}{\sum_{i\in S^{(k)}}\exp(\phi_u^\top\phi_I(i))} }{{\partial \phi_u}^2}\right]_{\phi_u=\hat{\phi}_{\mu,U}(u)}\:.
\end{align*}
Using the gradient we derived earlier in Appendix~\ref{app:gradient_of_logposterior}, we can equivalently write this as:
\begin{align}
    \mathbf{H} \commentout{&=-\left[\sum_{k=1}^{K}\frac{\partial\left(\phi_I(\rho^{(k)})-\frac{\sum_{i\in S^{(k)}}\exp(\phi_u^\top\phi_I(i))\phi_I(i)}{\sum_{i\in S^{(k)}}\exp(\phi_u^\top\phi_I(i))}\right)}{\partial \phi_u}\right]_{\phi_u=\hat{\phi}_{\mu,U}(u)}\nonumber\\
    &=\left[\sum_{k=1}^{K}\frac{\partial\left(\frac{\sum_{i\in S^{(k)}}\exp(\phi_u^\top\phi_I(i))\phi_I(i)}{\sum_{i\in S^{(k)}}\exp(\phi_u^\top\phi_I(i))}\right)}{\partial \phi_u}\right]_{\phi_u=\hat{\phi}_{\mu,U}(u)}\nonumber\\}
    &=\Bigg[\sum_{k=1}^{K}\bigg(
    \frac{\sum_{i\in S^{(k)}}\exp(\phi_u^\top\phi_I(i))\phi_I(i)\phi_I(i)^\top}{\sum_{i\in S^{(k)}}\exp(\phi_u^\top\phi_I(i))} - \nonumber\\
    &\frac{\sum\limits_{i\in S^{(k)}}\exp(\phi_u^\top\phi_I(i))\phi_I(i)}{\sum\limits_{i\in S^{(k)}}\exp(\phi_u^\top\phi_I(i))}\frac{\sum\limits_{i\in S^{(k)}}\exp(\phi_u^\top\phi_I(i))\phi_I(i)^\top}{\sum\limits_{i\in S^{(k)}}\exp(\phi_u^\top\phi_I(i))}
    \bigg)\Bigg]_{\phi_u=\hat{\phi}_{\mu,U}(u)}
\end{align}
}

\commentout{
\begin{table}[ht!]
  \centering
  {\footnotesize
  \begin{tabular}{|l||l|}
  \hline
  \textbf{PE Method} & \textbf{Top-10 Attributes}\\
    \hline
    Attribute EVOI & visually appealing, anime, depressing, tom hanks,\\ &betamax, thought provoking, action, japan, zombies, funny\\
    \hline
    Attribute Random & psychology, quirky, dvd-video, philosophical, alternate reality, \\ &french, beautiful, atmospheric, artificial intelligence, heist\\
    \hline
    IpA EVOI & anime, depressing, sci-fi, zombies, friendship, funny,\\ &chick flick, loneliness, philosophical, drugs\\
    \hline
    IpA Random & history, road trip, documentary, stephen king, \\
    &chick flick, gothic, space, dystopia, robert de niro, zombies\\
    \hline
 \end{tabular}
  } \\
   \caption{The Top-10 attributes in different PE methods that collects the most information. }
  \label{tab:top attr}
\end{table}
}
\commentout{
\section{Data Generation with RecSim NG}
\label{app:syntheticdata}

We use a stylized, but structurally realistic generative model to produce synthetic ratings and tag data for some of our experiments. Its purpose is twofold. First, it offers various parameters or ``knobs'' that can be used to generate data sets that can increase/decrease the level of difficulty faced by methods designed to extract the semantics of soft or hard attributes. Second, synthetic data generation provides us with a ``ground truth'' against we can test the effectiveness of our elicitation methods using soft attributes to improve recommendations made for users.

The generative process is given in the following stages: we first generate items (with their latent and soft attribute values); then users (with their utility functions); then user-item ratings; and finally user-item tags. The model reflects realistic characteristics such as item and user ``clustering,'' popularity bias, not-missing-at-random ratings, the sparsity of ratings, the relative sparsity of tags compared to ratings, etc.

Specifically each item $i$ is characterized by an \emph{attribute vector} $\bfv(i)\in [0,1]^D$, where $D = L+S$: $L$ dimensions correspond to latent item features and $S$ to soft attributes. For a soft dimension $L < s \leq L+S$, $v^s(i)$ captures the degree to which $i$ exhibits soft attribute $s$. We sample $m$ items from a mixture of $K$ $D$-dimensional Gaussian distributions (truncated on $[0,1]^D$) $\calN(\mu_k, \sigma_k)$, $k\leq K$, with mean vector $\mu_k \in [0,1]^D$ and (diagonal) covariance $\sigma_k$. We set $D$ to 25 and $K$ to 100 in our experiments. For simplicity, we assume all $\sigma_k$ are identical to 0.5, and sample means uniformly. Mixture weights are sampled uniformly at random and normalized.
For each item $i$, we also randomly generate a popularity bias $b_i$ from $[0,1]$ (whose purpose is described below).

Each user $u$ has a \emph{utility vector} $\bfw(u)\in [0,1]^D$ reflecting its utility for items. We sample $n$ users from the above $K$-mixture-of-Gaussian distribution similar to that for items. Here the means and variances of these distributions are same as the ones used for item distributions,
but the mixture weights are fully resampled. This step ensures that the generated samples of users and items are distributed in different parts of the latent ``topic space''.

Next, we generate the user-item ratings with the following steps:
\begin{enumerate}[(i)]
\item For each $u$, we draw $\NumR_u$ samples from a Zipf distribution with a power parameter $a = 1.05$ to reflect the natural power law over the number of ratings provided by users. 
We also set the maximum number of ratings by each user to $1,000$. 
\item To generate the candidate items to be rated by each user $u$, we
generate a set of $\Rated_u$ items, by sampling them without replacement from the overall set of items via a multinomial logit (or softmax) choice model, where the probability associated with each item $i$ is proportional to $e^{\tau \cdot (\bfw_u\bfv_i + b_i)}$. 
We set the temperature parameter $\tau$ to $1$ in our experiments.
\item For each user $u$ and item $i\in\Rated_u$, the rating $r_{ui}$ is generated as follows. We denote by $s(u,i):=\bfw_u\bfv_i + \veps$ be the score of item $i$, where $\veps$ is a small, zero-mean random noise. We then discretize all the scores provided by user $u$ into $5$ equally sized sub-intervals
assign a $1$ to $5$ rating to each item accordingly.
\end{enumerate}

For each soft attribute $s$ we assume there is a unique tag $g_s$ that users can apply when referring to that attribute. On the other hand, for each generic tag $g$, we denote by $s(g)$ the corresponding soft attribute (so $g = g_{s(g)}$). To complete the data-generation procedure we also generate user-item tags with the following steps. 
\begin{enumerate}[(i)]
\item For each user $u$, we generate $\PT_u$, the probability of tagging an item, from a mixture of two distributions: (a) a Dirac distribution at $0$ with weight $0<x<1$; and (b) a Uniform distribution
over $[p_-, p_+]$ with weight $1-x$. This reflects the fact that a large fraction of users never use tags, and among those who do, some users tag much more frequently than others. In our experiments, we set $x$ to 0.8, $p_-$ to 0.1, and $p_+$ to 0.5.

\item For each user $u$ we first generate the set $\Tagged_u$, which represents the items that are tagged by $u$. Here $\Tagged_u$ is a subset of rated items $\Rated_u$ such that each rated item will be tagged with (independent) probability $\PT_u$. This reflects the fact that a user will not tag an unrated movie, but may leave some rated movies untagged.\footnote{One could also allow, if desired, the propensity to tag to vary with the tag $g$, and/or bias the application of tags to higher-rated items.} For any item $i\not\in\Tagged_u$, the corresponding indicator value $t_{u,i,g} = 0$ for every tag $g$, which means that no tag is applied by user $u$ on item $i$.

\item For every (non-subjective) tag $g$, we use a user-independent threshold $\tau_g = 0.5$ indicating the degree to which an item must possess attribute $s(g)$ to be tagged with tag $g$ by a user. 

\item For every item $i \in\Tagged_u$ and tag $g$, we set the indicator value $t_{u,i,g} = 1$ (i.e., user $u$ applied tag $g$ to item $i$) if $v^{s(g)}(i) \geq \tau_g + \veps$ (where $\veps$ is a small, zero-mean random noise drawn independently from $\calN(0, 0.01)$ for each $(u,i,g)$). Otherwise the indicator value $t_{u,i,g}$ remains at $0$. 

\end{enumerate}
}



\end{document}